\documentclass[12pt]{iopart}

\usepackage[tight]{subfigure}
\usepackage{amssymb}
\usepackage{float}
\usepackage{graphicx}
\usepackage{color}
\newcommand{\lyxdot}{.}

\begin{document}
\title[Origin of Corrections to Mean-field at the Onset of Unjamming]{Using Point to Set Correlations to Probe the Unjamming Transition}  
\author{M Mailman$^1$ and B Chakraborty$^2$}
\address{$^1$Department of Physics, IPST and IREAP, University of Maryland, College Park, Maryland, 20742}
\address{$^2$Martin Fisher School of Physics, Brandeis University, Waltham, Massachusetts 02454 and }
\ead{\mailto{mailmm@brandeis.edu}, \mailto{bulbul@brandeis.edu}}

\date{\today}

\begin{abstract}
We present a detailed analysis of the unjamming transition in 2D frictionless disk packings using a static correlation function that has been widely used to study disordered systems.  We show that this point-to-set (PTS) correlation function exhibits a dominant length scale that diverges as the unjamming transition is approached through decompression.  In addition, we identify deviations from meanfield predictions, and present detailed analysis of the origin of non-meanfield behavior. A mean-field bulk-surface argument is reviewed.  Corrections to this argument are identified, which lead to a change in the functional form of the critical PTS boundary size $R_0$.  An entropic description of the origin of the correlations is presented, and simple rigidity assumptions are shown to predict the functional form of $R_0$ as a function of the pressure $P$.  
\end{abstract}\

\maketitle

\section{Introduction\label{sec:Introduction}}

It is widely believed that unjamming of disk packings is a singular
point: as $P\rightarrow0$ (or $\phi\rightarrow\phi_c$), a disk packing loses rigidity and ceases
to be a solid\cite{Jamming98,CoreyCompressionRoutine}. What is still up to debate is whether or not this point
represents a critical point in the sense of a true thermodynamic phase
transition. In critical phenomena, an important signature of a phase
transition is a characteristic length scale that emerges from the
statistical mechanical system and becomes the system size (diverges
in the thermodynamic limit) when the system reaches the critical point.
The origin of such a length scale lies in spatial correlations of the statistical
variables within individual microstates that become longer ranged as the critical point is approached.  As a result, such a length scale is an inherent property of static configurations.  

Measures of such a length scale in granular systems have been widely
sought after, and with some success. But, what is still missing is
an understanding of the correlations that are associated with the
relevant length scale in granular systems. We begin by discussing previous studies of the length scale in granular systems,
and their interpretation. This previous work includes the study of
correlations in velocities and non-affine displacements in unjammmed,
flowing granular packings under shear\cite{BarratHeussinger}, as
well as force fluctuations which are measured in response to point
perturbations in jammed packings\cite{EllenbroekLinResponse}. A theoretical
framework based on mean-field bulk-surface arguments is discussed,
which predicts a growing length scale\cite{MatthieuLengthScale,TkachenkoIsocounting}. 

In this paper, we expand on the work of \cite{jstatpaper}, where it was shown that a static correlation function
exhibits a length scale that diverges in the thermodynamic
limit for computer-generated packings of disks in 2D. This {}``Point-to-Set''
correlation function (PTS) is motivated by the Random First Order Transition
(RFOT) theory of glasses\cite{BiroliAdamGibbs,BiroliNPhys,BiroliReview,PTSLJ,PTSspin}. We calculate the PTS correlation function in 2D disk packings through
the use of the force network ensemble, which has been studied extensively
as a model of overcompressed jammed packings\cite{FNE,VanHeckeTigheReview,McNamara}.
We investigate the origins of this correlation function starting from an entropic
formulation of the space of solutions to mechanical equilibrium of the disk packings.
Through our analysis of the microscopic nature of the correlations, we gain insight into the unjamming transition, and especially,  the deviations
from the mean-field predictions.

\section{The Isostatic Argument\label{sec:The-Isostatic-Argument}}

The first indications that there might be a critical point associated
with jamming and unjamming of spherical grain packings dates back
to discussions of marginal rigidity by Maxwell\cite{MaxwellIso}, and has been revisited in for instance
\cite{TkachenkoIsocounting,MouzarkelInIsoStat}. In the limit of hard
spheres and an infinite system size in $d$ dimensions, the number
of contacts $z$ of a packing can be calculated exactly.  The hardness leads to a geometric
constraint: no two spheres can overlap. Therefore, for any two spheres
with diameters $D_{i}$ and $D_{j}$, the center-to-center distance
between the two spheres $R_{ij}\geq\left(D_{i}+D_{j}\right)/2$. The
coordinates that locate the centers of the spheres act as degrees
of freedom, with $dM$ such degrees of freedom for $M$ spheres, and
the overlap constraints are applied at each contact, of which there
are $M\left[z\right]/2$, if each grain has on average $\left[z\right]$
contacts. The $2$ takes into account the double-counting of contacts.
In order to satisfy each constraint and still be consistent, the number
of degrees of freedom must be greater than or equal to the number
of constraints: $dM\geq M\left[z\right]/2$. 

In the case of soft spheres at non-zero pressure, some spheres must
overlap. Instead, it is the contact forces that are determined by
the constraints imposed by mechanical equilibrium. There are $M\left[z\right]/2$
contact forces, and $dM$ mechanical equilibrium equations which must
be satisfied. Therefore, $dM\leq M\left[z\right]/2$ in order for
a packing of soft spheres to be mechanically stable.

In the limit of hard spheres and zero pressure, both inequalities
can be satisfied only when both inequalities are equalities: $dM=M\left[z\right]/2$
so that $\left[z\right]=2d=z_{0}$. In particular, for two dimensions
the {}``isostatic'' value $z_{0}$ is 4. 

When grains are soft, $\left[z\right]$ can become greater than $z_{0}$
as the packing fraction is increased. With respect to unjamming of
soft grains, the observation that there is a minimal value of $\left[z\right]$
below which the pressure of the packing must go to zero and cannot
be mechanically stable suggests that the isostatic point $z_{0}=2d$
might be a critical point, since states with $\left[z\right] > z_{0}$ are expected to flow to a jammed, mechanically stable fixed point and state with $\left[z\right] < z_{0}$ to an unjammed, gas-like state. This transition, however, happens out of equilibrium, and an interesting question to ask is whether there is a diverging length scale associated with the isostatic point.

\section{The Hunt for a Granular Length Scale}

\subsection*{Bulk-Surface Argument and the Isostatic Length Scale \label{sub:Bulk-Surface-Argument}}

A length scale in overcompressed jammed packings that increases as they are decompressed 
(pressure $P\rightarrow0$) is predicted from a bulk-surface argument.  
Based on counting arguments that date back to Maxwell\cite{MaxwellIso}, and more recently discussed in \cite{TkachenkoIsocounting,MouzarkelInIsoStat}, one begins with 
$M\left[z\right]/2$ force bearing contacts in a subregion
of a packing, where $M$ is the number of grains in the subregion
and $\left[z\right]$ is the average number of contacts per grain
of the packing.    For each of the $M$ grains, there are
$dM$ equations of mechanical equilibrium (ME) constraining the forces
at each of the $M\left[z\right]/2$ contacts (in $d$ dimensions).  As is discussed below, in the limit of very stiff grains, the deformations of two grains in contact are impossible to resolve even though contacts can carry a great deal of force.  The only constraints on these contact forces, then, are ME equations, and force laws can be ignored.  There are $M\left[z\right]/2-dM=M\delta z/2$ unconstrained contact force magnitudes, which are taken to be the degrees of freedom, with $\delta z = \left[z\right] - z_0$.  Finally, if one defines a subregion by a set of boundary grains, these
grains provide an additional $A$ constraints from the boundary. Roughly,
$A\propto\sqrt{M}$ in 2D. The number of grains can be estimated using
the definition of the packing fraction: 

\begin{equation}
\phi=c \frac{M r^{2}}{L^{2}}\rightarrow M=\frac{\phi}{c}R^{2}\label{eq:CoarseGraining}
\end{equation}

The unitless length $R$ is the ratio of size of the subregion
$L$ to the radius $r$ grains (for our purposes, systems are bidisperse and $r$ is taken to be the diameter of the smaller grain), and the parameter $c$ characterizes the polydispersity.

Using \ref{eq:CoarseGraining}, the number of excess contact force degrees
of freedom $\delta n$ can be written as a function of $R$. In 2D,

\begin{equation}
\delta n_{nom}=\frac{M\delta z}{2}-A=\alpha\phi R^{2}\delta z-\beta\sqrt{\phi}R\label{eq:continuumBulkSurface}
\end{equation}

\noindent where $\alpha$ and
$\beta$ are constants.  This is a nominal expression for $\delta n$, hence the subscript {}``nom," and will be discussed in more detail in the proceeding sections.

At this point, it's worth contrasting the bulk-surface argument presented
here with those found in the literature \cite{MatthieuLengthScale,TkachenkoIsocounting}.
In previous constructions of the bulk-surface argument, the boundary
of the subregion is said to contribute a set of fixed, {}``frozen''
contact forces, so that the total number of excess contacts $M\delta z/2$
is ${\it {\it reduced}}$ by $A$ contacts. This should be contrasted
with the above construction, where the boundary term contributes ${\it additional}$
boundary constraints on the variable contact forces. Near isostaticity, there is no difference between the two constructions,
but away from isostaticity, the boundary term in the formulation of \cite{MatthieuLengthScale,TkachenkoIsocounting}, when contributing extra
frozen contacts, should scale with $z$: $A\propto\sqrt{M}\left[ z \right]$, since
on average there are more contacts on the boundary if there are more
contacts in the packing as a whole. 

Frozen contact forces differ from additional constraints due to ME: the former reduce the number of degrees of freedom \cite{MatthieuLengthScale,TkachenkoIsocounting} and are proportional in number to $\left[z\right]$ while the latter serve to further determine the existing contact forces and are proportional in number to $M$.  An important length scale is determined by  $\delta n_{nom}\left(R_{0}\right)=0$.  This length scale defines the size of the smallest region within which the mechanical equilibrium equations can be satisfied, and 
$R_{0}\propto\delta z^{-1}$ if one assumes that the dependence
on $\phi$ is small.  In the existing literature \cite{MatthieuLengthScale},
this length scale, referred to as the isostatic length scale
$l^{*}$, is deduced from a form of the  bulk-surface argument that depends
on frozen boundary contact forces.  There is no difference in the results of the two constructions as $\delta z\rightarrow0$. However, with boundary contact
forces frozen, $l^{*} \sim \frac{\left[z\right]}{\delta z}$, and can differ significantly from $R_0$ at large overcompressions. In \cite{EllenbroekLinResponse} the isostatic length scale is measured indirectly
through the response to a point-force perturbation to mechanically stable
overcompressed packings, and is shown to scales as $1/\delta z$.  In later sections, it will be made clear why a bulk-surface
argument of the form \ref{eq:continuumBulkSurface} is preferable
in the context of the PTS correlation.

The vibrational spectrum of granular packings as defined by the dynamical
matrix provides
more insight into the nature of unjamming. The excess degrees of freedom
$\delta z$ can be related to the pressure of the packing through
the vibrational density of states\cite{MatthieuLengthScale,SilbertLiuNagel}.
A critical frequency $\omega^{*}$ is associated with the highest
energy debye-like mode of a marginally stable packing. Anomalous modes
(modes whose density of states are roughly independent of the frequency)
appear at energies higher than $\omega^{*2}$, and this energy difference
is proportional to $-P$. So, the energy change due to an imposed
pressure $P$ is $\Delta E\leq\omega^{*2}-AP$ and the critical frequency
is $\omega^{*}\propto P^{1/2}$. To relate $\omega^{*}$ to the contact
number, one must look at the characteristic size of the anomalous
modes. It turns out that anomalous modes with $\omega \ge \omega^*$ are quasi-localized with localization length $\ge R_{0}$\cite{ZoranaQuasiLocal}.  Assuming a linear dispersion relation $\omega\propto ck$, the critical frequency is: $\omega^{*}\propto\frac{1}{R_{0}}\propto\delta z$, and therefore, $\delta z\propto P^{1/2}$\cite{WyartThesis,EllenbroekThesis}. 

It should be noted that the validity of the dynamical matrix approach
to deriving the vibrational spectrum has been recently called into question\cite{Carlnonharmonic}.
Numerical studies show that in the limit of infinite system sizes,
arbitrarily small perturbation amplitudes lead to contact breaking,
and mixing of the eigenmodes of the dynamical matrix\cite{Carlnonharmonic}. This result calls into question
any analysis of response of jammed packings based on the vibrational density of
states. While the result $\delta z\propto P^{1/2}$ is well established
in simulations (we have verified that our numerics reproduce this result), the justification of
this relationship relies on the existence of anomalous modes in the vibrational spectrum. Furthermore,
the explanation that the length scale measured in collective response is associated with the extent
of such anomalous modes becomes questionable, further motivating our choice to look for a static measure of a correlation length.

There are two aspects to the bulk-surface argument which are mean-field
in nature. The first is the relation between $P$ and $\delta z$.
In principle, $\Delta E$ should depend on factors which contain the
force vector orientations, but it is assumed that these vectors are
randomly distributed in the packing so that their effects on $\Delta E$
average to a constant. If, on the other hand, the local configuration
of force vectors depended heavily on the orientations of nearby force
vectors, as one might expect for instance simply because of the overlap
constraints, this might not be a reasonable assumption. It turns out
that at least for isotropically compressed packings, the packing geometry
is sufficiently disordered that this assumption is sound\cite{MatthieuLengthScale}.

The second mean-field assumption is built into the counting that goes into the bulk-surface argument. The assumption that the number of excess force degrees of freedom is given by the difference between
the number of force variables and the number of ME equations requires
that these equations are all independent, since a linear system has
a number of undetermined variables equal to the difference between
the total number of variables and the number of $\emph{independent}$
equations. As will be detailed in the sections below, this assumption
does fail significantly as $P\rightarrow0$. 

\subsection*{Stress Correlations}

The work of \cite{BarratHeussinger} on transverse grain displacement correlations
in simulations of 2D disk packings under quasistatic shear strain reports a growing length
scale for packing fractions below $\phi_{J}$. Interestingly,  they find no evidence for a growing length scale as 
$\phi_{J}$ is approached from above.  These results  are consistent with continuum
elastic theory \cite{Maloney} which shows that shear-strained packings
can be thought of as random local forces applied to a homogeneous
elastic sheet. Because of this homogeneity, the system size becomes
the only relevant length scale of the system. Furthermore, in \cite{Lois2009}
a field-theoretic calculation of pressure fluctuation correlations
for both isotropically compressed and sheared systems shows that such
correlations exhibit a power-law behavior, with the only length scale
emerging as the grain size, independent of the pressure. 

At this point, it would seem that (2-point) stress correlations do
not exhibit a growing length scale. Neither do displacement correlations,
even though force fluctuations seem to scale with the isostatic length
scale $l^{*}$. The latter requires one to consider the response to a point perturbation, and unlike systems in thermal equilibrium, the relation between response and correlations  is not established in jamming systems.  The question of whether there is a static correlation function that can be identified with the unjamming transition is, therefore, still wide open. 

The importance of probing a static correlation function
can be understood from contrasting the results from the non-equilibrium unjamming transition to that of, say,
a simple Ising model: for a large enough system, a single spin configuration
of the magnet equilibrated at some temperature $T$ will exhibit a
spin-spin correlation length that grows as $T_{c}$ is approached.
One does not have to consider properties of the dynamics which brings
one spin configuration into another. If the unjamming transition is
to be thought of as a critical point and not a kinetic freezing transition\cite{ChandlerRef}, 
a static correlation function must show a diverging length scale.
In section \ref{sec:Point-to-Set-Correlations}
a new type of correlation function is discussed, which is specifically
designed to overcome the added complexity of amorphous systems where
critical behavior is still to be expected.

\section{Point-to-Set Correlations \label{sec:Point-to-Set-Correlations}}

In the physics of the glass transition, the question of a static correlation
function associated with the arrest of dynamics and the onset
of the glassy state is also very important, and has been investigated extensively\cite{BiroliReview}.  There has been speculation that the jamming transition is the zero temperature
limit of a glass transition driven by compression or shear \cite{Jamming98}.  Regardless, in both glassy and granular systems there exist metastable states which characterize the statistical mechanics of the glassy or jammed phases \cite{BouchaudIdeas}, and ideas from the glass literature can be used to probe the existence of a growing static length scale associated with unjamming.

In the Random First Order Transition (RFOT) theory of the glass transition,
a correlation function is defined\cite{BiroliAdamGibbs,BiroliReview}, 
which can probe growing amorphous order. In phase transitions characterized by a well-defined order parameter, a 2-point correlation
function is sufficient to distinguish between the disordered and ordered
state\cite{HuangStatMech}. 
However, pre-existing knowledge of the structure of the ordered state
is built into the definition of the two point correlation function.  A simple example demonstrating this is frustrated magnets, where
the 2-point spin-spin correlation function fails to clearly identify
the correlations. Instead, one engineers a correlation function which
is sensitive to an {}``ordered state'' of staggered spins. This correlation function is constructed with ${\it a\, priori}$
knowledge of the ordered state.  For an amorphous system, where the
free energy landscape can become quite complex, with minima corresponding
to different configurations of the particles, no such ${\it a\, priori}$ knowledge exists.  

Without an intuition for the relevant symmetry needed to describe the jammed or glassy phase to motivate the construction of a correlation function, one instead begins with the {}``definition'' of a thermodynamic second
order phase transition\cite{BiroliNPhys}: the influence of boundary
conditions on the bulk of the system grows as the critical point is
approached. Imagine that an Ising magnet is equilibrated at some $T$
and used to define a boundary condition of spins that are characteristic
of that temperature. Then a configuration of spins of size $R$ is
allowed to re-eqiulibrate with respect to that particular boundary
condition. If $R<\xi\left(T\right)$, the resulting spin configuration
will not change much from that of the original equilibrated magnet.
However, if $R>\xi\left(T\right)$ then the configuration of spins
that is equilibrated with respect to the boundary conditions acts
essentially as a free set of spins, insensitive to the boundary conditions,
and can re-equilibrate to a very different configuration. An appropriately
defined overlap of the equilibrated magnet and the magnet with respect
to the boundary conditions captures the relevant correlations and
exhibits the diverging length scale $\xi\left(T\right)$. Such a correlation
function is known as the Point-to-Set (PTS) correlation function\cite{PTSspin},
since it captures the correlations of a single degree of freedom such
as a spin with a set of boundary degrees of freedom that are fixed
as boundary conditions.

The advantage of this correlation function over typical 2-point correlation
function is that it generalizes easily to amorphous systems where
the degrees of freedom can be more complicated but the boundaries
are still easily defined. For instance, in \cite{PTSspin} the PTS
correlation function for the p-spin system on a Bethe lattice is calculated
exactly and shown to exhibit a growing length scale as the critical
temperature is approached, even though the spin configurations of
such systems are amorphous. In a more realistic Lennard-Jones glass
forming liquid\cite{PTSLJ}, a PTS correlation function is shown to
exhibit correlations that persist for larger values of $R$ as the
temperature is decreased. 

With the PTS correlation function, there is no general framework for
the definition of the overlap. It appears to be very system specific.
For instance, in \cite{PTSLJ} the overlap of the {}``reference state''
(the equilibrated configuration of particles) with the {}``pinned
state'' (equilibrated with respect to a fixed boundary) is defined
as the product of the thermally averaged occupation number of discrete cells which are used to discretize space.

\begin{figure}[t]
\begin{centering}
\includegraphics{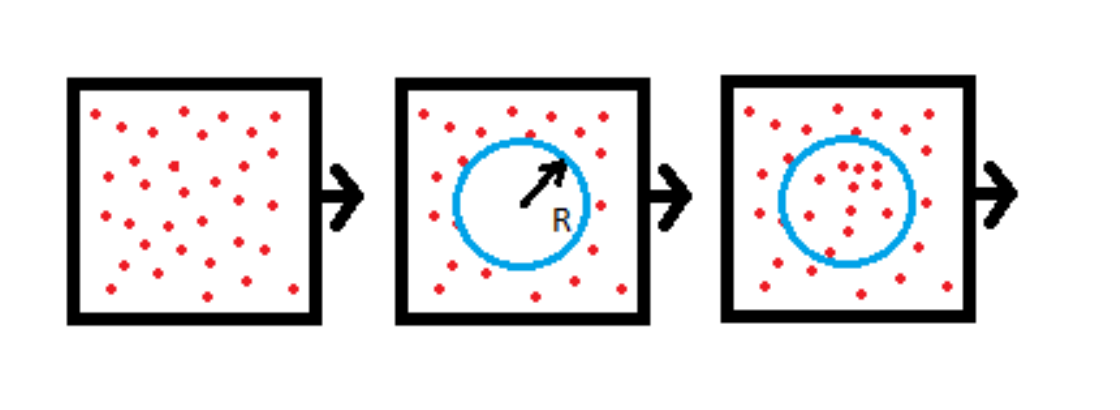}
\par\end{centering}

\caption{In general, the protocol for calculating a PTS correlation function
involves finding an equilibrated state of the system (left, with red
dots representing degrees of freedom). Then, a boundary of size $R$
is defined, where degrees of freedom outside are kept fixed (middle)
and inside are allowed to continue to fluctuate, but with respect
to the boundary conditions imposed by the fixed degrees of freedom.\label{fig:schematic_of_PTS}}

\end{figure}

Roughly speaking, what we learn from these examples is that a PTS
correlation function should have the property that it compares a typical
{}``reference'' configuration of the system to a {}``pinned''
configuration of the system which is typical given a particular set
of boundary conditions, themselves derived from the reference configuration.
Some overlap must be defined that captures how similar the reference
and pinned states are. The dependence on distance in the correlation
function always results from the size of the bounded region of the
pinned state (figure \ref{fig:schematic_of_PTS}).

\section{The Force Network Ensemble and a Granular PTS\label{sec:The-Force-Network_Ensemble_Granular-PTS}}

\subsection*{Definition of the Force Network Ensemble}

The Force Network Ensemble (FNE) has been studied extensively as a
statistical mechanical model of granular matter\cite{VanHeckeTigheReview}.
While there are several variations on the model studied in the literature\cite{WheelMove,McNamara,FNE},
the key assumption made when developing the FNE is that when grains
are sufficiently rigid, infinitesimally small deformations of grains
(or, simply deformations that are too small to observe experimentally)
lead to significant fluctuations in the contact forces. With this
assumption, the force law coupling grain deformations to forces is
no longer relevant. For a given fixed configuration of mechanically
stable grains (referred to here as an MS), the contact forces can
take on any positive values that do not violate the constraints of
mechanical equilibrium (ME). When the grains are frictionless, the
forces $\vec{f}_{ij}$ always lie along the vector normal to the point
of contact $\hat{r}_{ij}$, which in the case of disks is the center-to-center
separation vector between grain $i$ and its neighbor $j$. The ME
constraints are linear in the contact force magnitude, and are expressed
as\cite{NewtonsLaws}: 

\begin{equation}
\sum_{j=1}^{z_{i}}\hat{r}_{ij}f_{ij}=0\label{eq:newtons law}\end{equation}

\noindent In addition, there are generally global constraints from the force moment-tensor of $M$ grains 

\begin{equation}
sigma_{\alpha\beta}=\sum_{i=1}^{M}\sum_{j>i}^{M}\frac{f_{ij}r_{ij}^{\alpha}r_{ij}^{\beta}}{ r_{ij}}\label{eq:stresss-tensor}
\end{equation}

\noindent (for $\alpha,\beta=x,y$ in 2D) on the force
networks, which are linear in the contact forces. These global constraints
are inhomogenous since the stress of a jammed packing is greater than
zero. The linear system of equations is expressed as a matrix equation

\begin{equation}
A\vec{f}=\vec{b}
\label{eq:LinearSystem}
\end{equation}

The matrix $A$ is made up of the geometric information of the MS.
In particular, it contains the components of $\hat{r}_{ij}$. The
vector $\vec{f}$ is a list of contact force magnitudes $N$ long,
where $N$ is the total number of contacts for the MS. The vector
$\vec{f}$ represents a particular force ${\it network}$ that satisfies
ME for a given MS, and should not be confused with a force vector
$\vec{f}_{ij}$ that is applied to a particular grain. The vector
$\vec{b}$ is a list of length $2M+3$ (in 2D) that contains the values
of the constraints. Entries which correspond to force balance equations
are zero, while the entries corresponding to components of the stress
tensor are non-zero constants: $\vec{b}=\left(0,0\ldots\sigma_{xx},\sigma_{xy,}\sigma_{yy}\right)$. 

The matrix $A$ of the linear system in Eq. \ref{eq:LinearSystem}
is rectangular with dimensions $2M+3$ by $N$ in 2D. A packing
with periodic boundary conditions will result in two more constraints
on the contact forces in 2D\cite{VanHeckeTigheReview}. The istostatic
argument of section \ref{sec:The-Isostatic-Argument} is easily expressed
in terms of the shape of $A$: assuming that the equations of ME are
all ${\it independent}$ (an assumption that will be discussed further
in the proceeding sections) the packing is isostatic, and hence the
linear system \ref{eq:LinearSystem} is precisely ${\it determined}$,
if A is square. If $A$ is rectangular with more rows than columns,
the packing is hypostatic, and therefore Eq.\ref{eq:LinearSystem}
${\it overdetermined}$, and if $A$ is rectangular with more columns
than rows, the packing is hyperstatic and therefore Eq.\ref{eq:LinearSystem}
is ${\it underdetermined}$. For the case where Eq.\ref{eq:LinearSystem}
is underdetermined, there is an infinite set of force networks $\left\{ \vec{f}\right\} $
that satisfy ME for the given MS. This set makes up a force network
ensemble for the given MS, referred to as the MS-FNE. It's important
to notice that two MS-FNE's are not interchangeable; a set of force
networks for a given geometry is not valid for any other geometry. 

Sampling force networks for an amorphous geometry
requires solving Eq.\ref{eq:LinearSystem}. When the linear system
is underdetermined, the matrix $A$ has a nullity greater than zero,
and so the solutions of $A\vec{f}=\vec{0}$ are spanned by null space
of dimension $\delta z$ (when $A$ is a full rank matrix). The singular
value decomposition (SVD) of $A$ results in the basis $\left\{ \hat{g}\right\} $
of the null space of $A$. Since Eq.\ref{eq:LinearSystem} is inhomogeneous,
$\left\{ \hat{g}\right\} $ are not solutions to Eq.\ref{eq:LinearSystem}.
Actually, $\left\{ \hat{g}\right\} $ are solutions at exactly zero
pressure, which requires each basis vector to have both positive and
negative elements, violating the positivity constraint. The homogeneous
solutions $\left\{ \hat{g}\right\} $ must be added to a particular
solution $\vec{f}_{0}$ which satisfies ME for the given MS and which
satisfies the inhomogeneous constraints, including the components
of the stress tensor. A good choice of $\vec{f}_{0}$ is easily found
by applying a force law, for instance linear spring interactions which
are used throughout this work, to construct a given MS and then extract a force
network from the geometry. In summary, for all $\left\{ \hat{g}\right\} $
such that $A\hat{g}=0$, there is a solution $\vec{f}=\vec{f}_{0}+c\hat{g}$
to Eq.\ref{eq:LinearSystem} for an amplitude $c$.

A method for sampling force networks for amorphous geometries then
amounts to finding a particular MS using any packing protocol, solving
for the null space of $A$ using numerical SVD routines, and choosing
amplitudes $c$ at random to construct new force networks $\vec{f}$
from $\vec{f}_{0}$ and $\left\{ \hat{g}\right\} $. As with the wheel
move\cite{WheelMove}, which can be shown to be a particular null vector of the $A$
corresponding to the triangular lattice, the positivity constraint
must be obeyed, and so any $\vec{f}=\vec{f}_{0}+c\hat{g}$ must be
regected if the resulting $\vec{f}$ has any negative elements. This
approach to sampling the FNE was first developed in \cite{McNamara}.

\subsection*{Application of a Frozen Boundary}

\begin{figure}[t]
\begin{centering}
\includegraphics[scale=0.5]{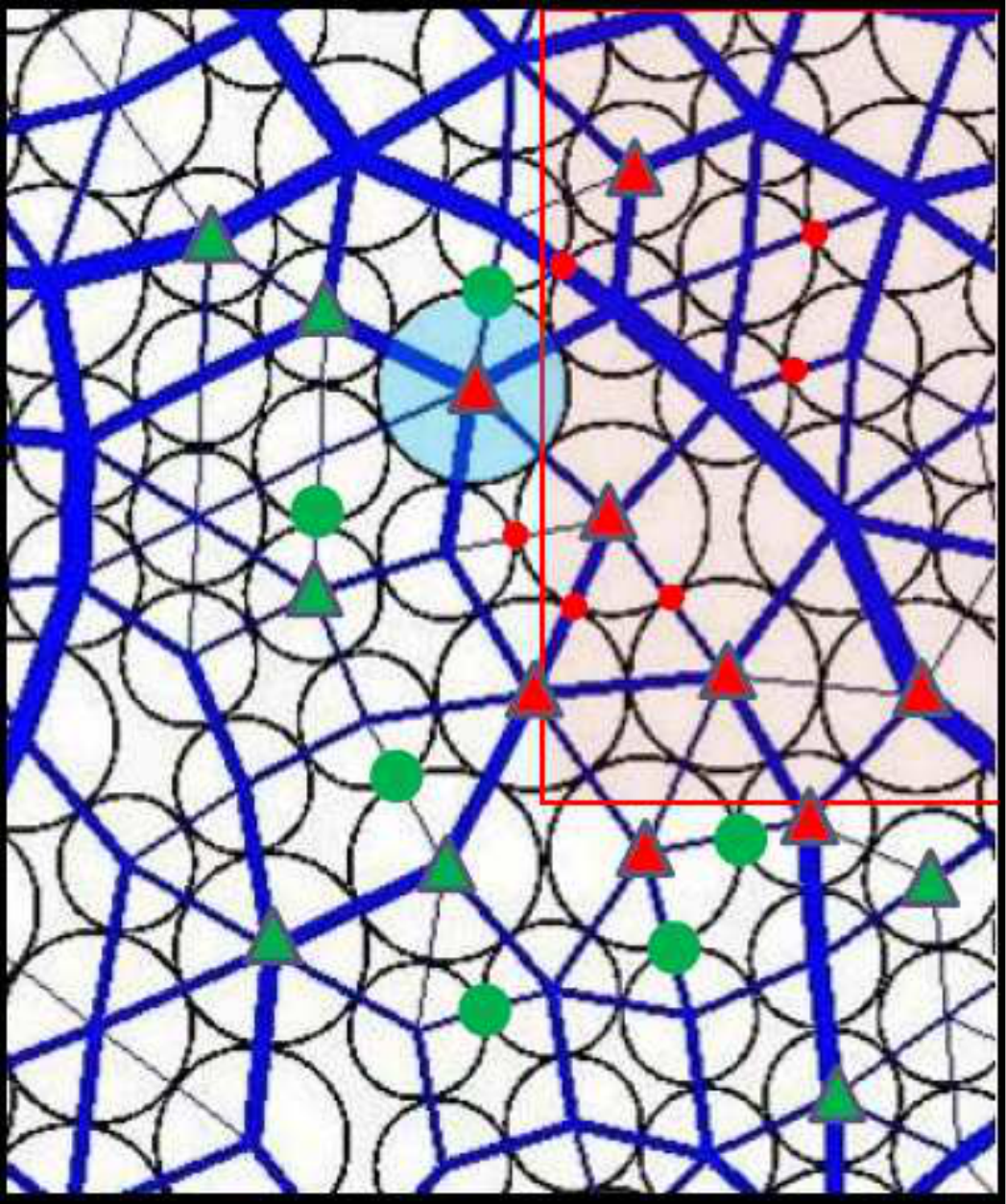}
\par\end{centering}

\caption{A sample disk packing with a boundary drawn in red, illustrating the
effects of the boundary on $A\left(R\right)$. Red triangles identify
grains which contribute ME equations to $A\left(R\right)\vec{f}\left(R\right)=\vec{b}\left(R\right)$
, and red dots are variable contact forces. Green circles, on the
other hand, are not considered variable contact forces, and green
triangles identify grains which do not contribute extra constraints
to the linear system. The blue grain is an example of a grain that
plays the special role of contributing ME equations to the linear
system, but which only has two fluctuating contact forces (the contact
forces that cross the boundary). The rule is that if the center of
a grain lies within the boundary, all of its contact forces will be
variables. \label{fig:definition_of_boundary}}

\end{figure}

It is not yet clear how to construct the PTS correlation function
for granular systems. A procedure analogous to that of LJ glass formers
might be to remove grains from a subregion of a packing and replace
them with a configuration of grains that originates from another packing,
and energy minimize to find the new mechanically stable packing while
keeping the boundary grains fixed. This approach presents several
issues. First, without thermal fluctuations it is not clear that the
final state is in equilibrium. There are large energies associated
with grain overlaps at the boundary. More fundamental though is the
problem of defining the correlation function. Since relevant variables
such as local stresses and forces are generally defined with respect
to the individual grains or contacts, and number of grains and contacts
can vary between subregions (as well as their locations), there is
no clearly defined overlap. With the added machinery of the MS-FNE,
though, a well-defined correlation function presents itself. For a
given MS, we define a PTS correlation function probes the correlations between
valid force networks that make up the FNE. The PTS correlation function
$C$ is then well defined as the inner product of any valid force
network $\vec{f}$ with the initial force network $\vec{f}_{0}$,
normalized by the magnitudes of the force networks to guarantee that
$C$ is never greater than 1: $C=\hat{f}_{0}\cdot\hat{f}$. But, this
correlation function still is not a function of a boundary size $R$.
Fortunately, the FNE allows for the ${\it net}$ force on a particular
grain to be fixed to a value other than zero through the addition
of inhomogeneous constraints to $A$. First, a boundary of size $R$
is defined on a subregion of the packing. For the work discussed here,
the boundaries are always square to match the symmetry of the simulation
box. Grains that are on the interior of the boundary are considered
in force equilibrium. They contribute ME equations to $A$ which are
homogeneous. Grains on the boundary, on the other hand, contribute
inhomogeneous constraints to $A$; the neighbors of those boundary
grains that are exterior to the boundary exert a fixed net force on
the boundary grain. These exterior grains do not themselves contribute
ME equations to $A$. All of the contact forces which are associated
with grains that have centers which lie interior to the boundary are
considered {}``free'' and are allowed to fluctuate. Contacts outside
of the boundary are not allowed to fluctuate and do not contribute
to the columns of $A$ (see figure \ref{fig:definition_of_boundary}).
The matrix $A$ now depends on the boundary size $R$. As $R$ increases,
more grains and contact forces fall into the interior of the boundary,
and so the dimensions of $A$ grow. In addition, the vector $\vec{b}$,
which now depends on $R$ as well, will have entries corresponding
to the fixed effective external forces being exerted on the boundary
grains from their exterior neighbors. The matrix equation $A\left(R\right)\vec{f}\left(R\right)=\vec{b}\left(R\right)$
must now be solved using the approach outlined for sampling the MS-FNE.
A well defined PTS correlation function can now be constructed so
that it is dependent on the size of the boundary: $C\left(R\right)=\hat{f}_{0}\cdot\hat{f}\left(R\right)$. 

As was discussed in section \ref{sub:Bulk-Surface-Argument}, the
bulk-surface argument is chosen to have a boundary term which corresponds
to the number of grains contributing ME equations to the constraints
on the contact forces, rather than a count of the number of {}``frozen''
contact forces. Here, a bulk-surface argument based on frozen contact
forces would be an error. There is no way of fixing the contact forces
at the boundary. Only ${\it net}$ ${\it forces}$ on each boundary
grain can be fixed. Fixing contact forces would involve fixing elements
of $\vec{f}$, which cannot be done given the approach to sampling
the FNE described here, since there is no control over the particular
values that the members of $\left\{ \hat{g}\right\} $ can take when
computing the SVD of $A$. 

The correlation function $C\left(R\right)$ is computed by averaging  over many
force networks $\vec{f}\left(R\right)$ and, once a sufficient sampling
of the FNE has been completed, the sampling is repeated for many MS.
Averages over the FNE are identified with $\left\langle \ \right\rangle $
brackets, and quantities that are averaged over the set of MS are
identified with $\left\langle \ \right\rangle _{g}$ brackets. Also,
the size of the bounding box $R$ can be varied and the process repeated
for different values of $R$. The only relevant microscopic scale
is the grain diameter $D$, so in practice the parameter that is controlled
is the scaled boundary size $R=R/D$. In the end, the correlation
function studied here for disk packings is $\left\langle \left\langle C\left(R\right)\right\rangle \right\rangle _{g}=\left\langle \left\langle
\hat{f}_{0}\cdot\hat{f}\left(R\right)\right\rangle \right\rangle _{g}$.

\section{Numerical Results for PTS }

\subsection*{The Correlation Function $C\left(R\right)$ \label{sub:The-Correlation-Function}}

\begin{figure}[t]
\subfigure[]{\includegraphics[scale=0.53]{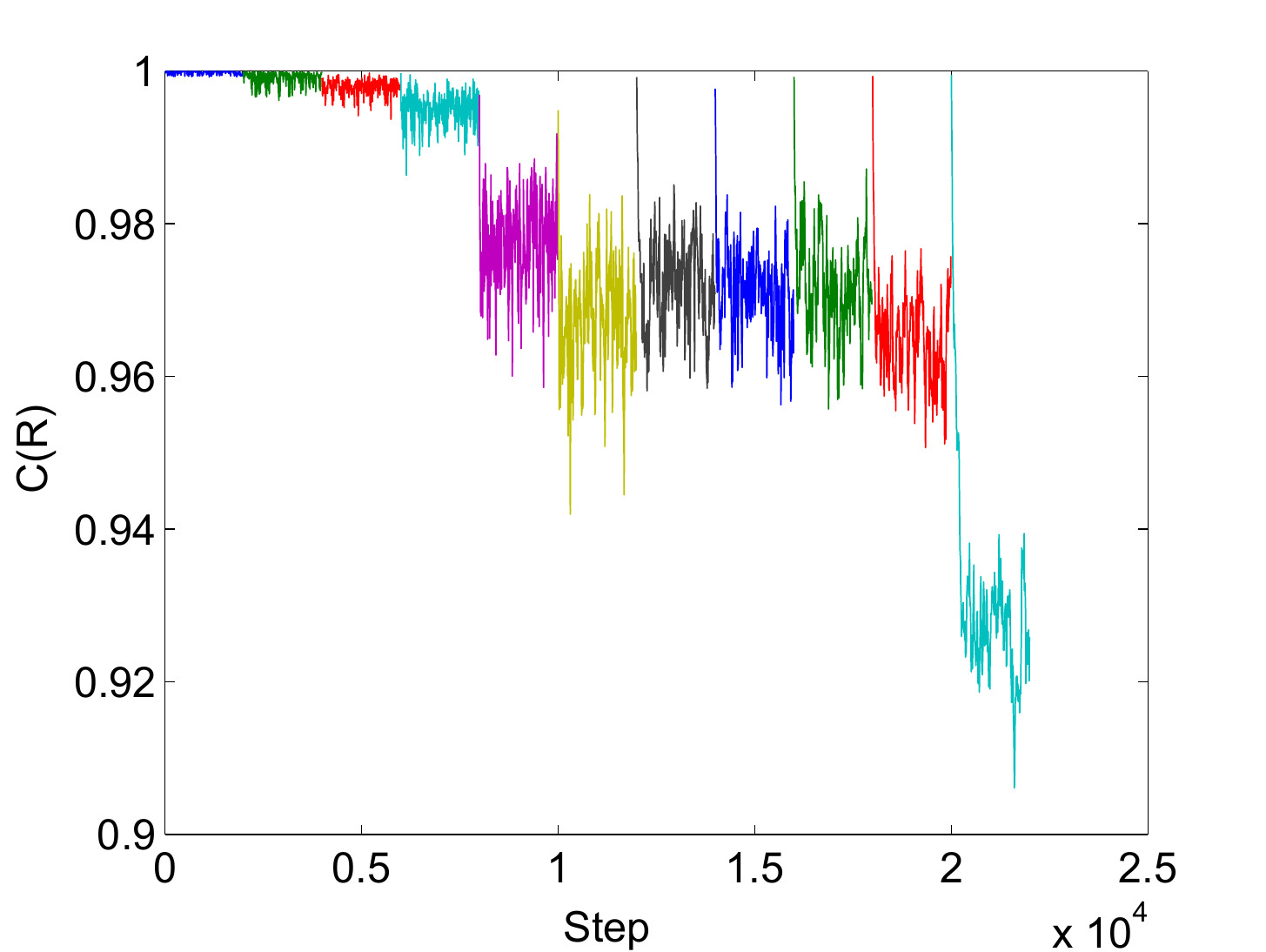}

}\hfill{}\subfigure[]{

\includegraphics[scale=0.53]{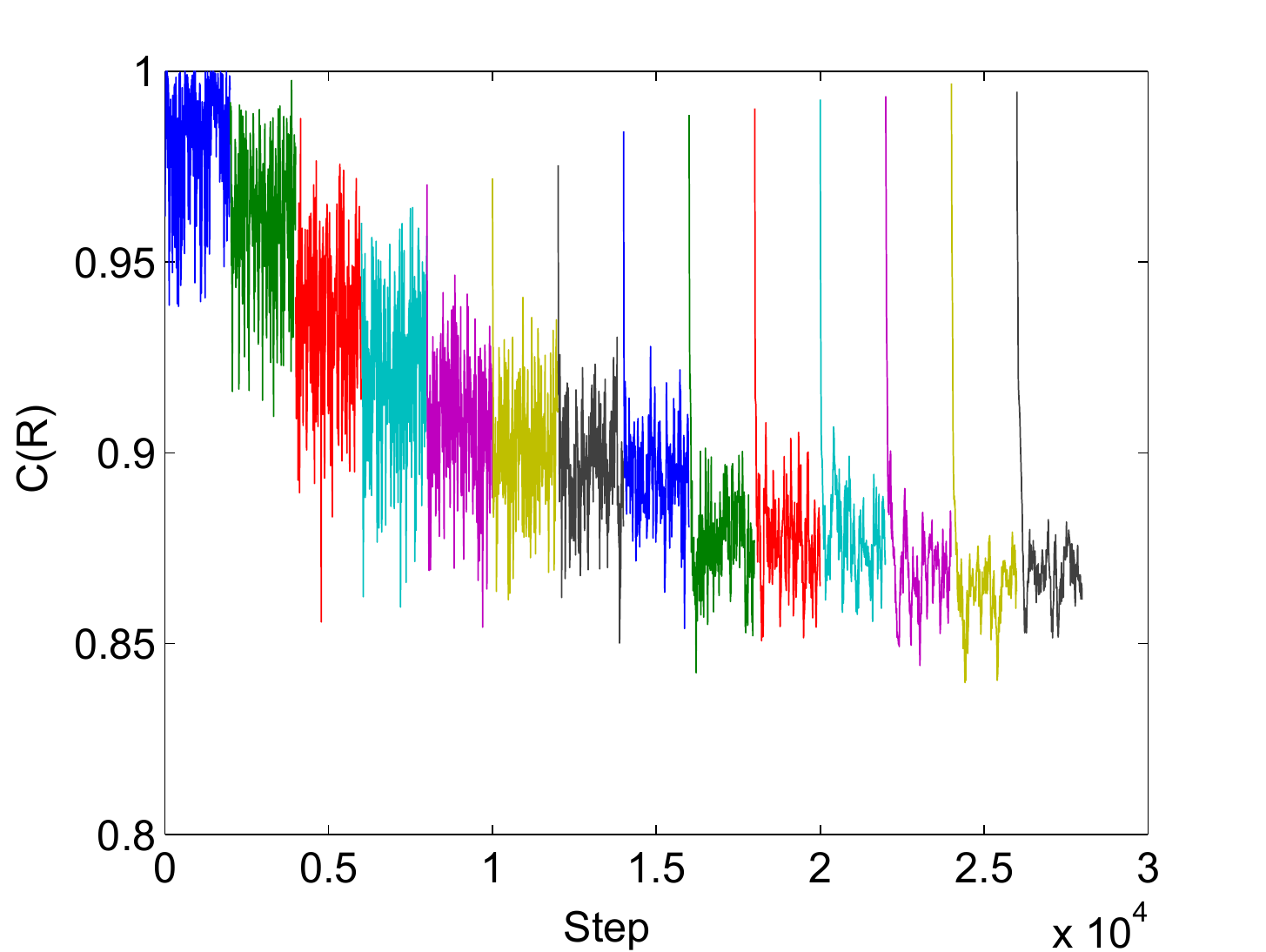}

}

\caption{(a) A single trajectory for $C\left(R\right)$ at an overcompression
of $\delta\phi=0.01$. Changes in the color represent an increase
in $R$, after which the sampling is allowed to re-equilibrate
to the new boundary conditions. The changes in $C\left(R\right)$
with $R$ are abrupt, but for a particular value of $R$, $C\left(R\right)$
quickly decays and then fluctuates about an equilibrium value. (b)
The same, but for $\delta\phi=0.1$. \label{fig:equilibrium test}}
\end{figure}

For systems of disk packings ranging from 30 to 900 grains in 2D,
bidisperse with one third of the grains 1.4 times larger than the
other two thirds\cite{bidisp1,bidisp2}, $\left\langle \left\langle C\left(R\right)\right\rangle \right\rangle _{g}$
has been measured. The MS averaging is done over 40 packings. The
FNE averaging is done over $10^{6}$ different forces networks. Force
networks are found by creating a high dimensional random walk in the
null space of solutions to $A\left(R\right)\vec{f}\left(R\right)=\vec{b}\left(R\right)$,
with a random step size chosen from a uniform distribution on an interval
$\left[-c,c\right]$ where $c$ is some fixed constant. The random
walk always begins from the initial force network $\vec{f}_{0}$.
In practice, a value of $c=0.05$ is found to keep the success rate
of the sampling high (as few steps as possible violate the positivity
constraint) while quickly moving away from the initial force network
$\vec{f}_{0}$ for a wide range of overcompressions and system sizes.
For the largest system size, figure \ref{fig:equilibrium test} illustrates
the trajectory of the random walk by calculating $C\left(R\right)=\hat{f}_{0}\cdot\hat{f}_{i}\left(R\right)$
for each random walk step $i$. When $R$ is changed, value of
$C\left(R\right)=\hat{f}_{0}\cdot\hat{f}_{i}\left(R\right)$
quickly drops and fluctuates around some value for that $R$. This
acts as a direct verification in terms of the correlation function
that the force network sampling has been allowed enough time to {}``equilibrate,''
and that the correlation function is measuring equilibrium properties
of the FNE. 

The results for $\left\langle \left\langle C\left(R\right)\right\rangle \right\rangle _{g}$
for the largest system size of 900 grains is shown in \cite{jstatpaper}.
At small $R$, the correlation function is nearly 1, and at some
value of $R$ it begins to decay. As will be discussed below, the
tail is a power law in $R$, and does not produce a length scale.
However, the crossover value of $R$ where the correlation function
begins to decay does exhibit the characteristics of a critical length
scale. This crossover value is referred to here as $R_{0}$ because
of its relationship to the $R_{0}$ of the bulk-surface argument,
which will be discussed in more detail below.

\begin{figure}[t]
\begin{centering}
\includegraphics[scale=0.53]{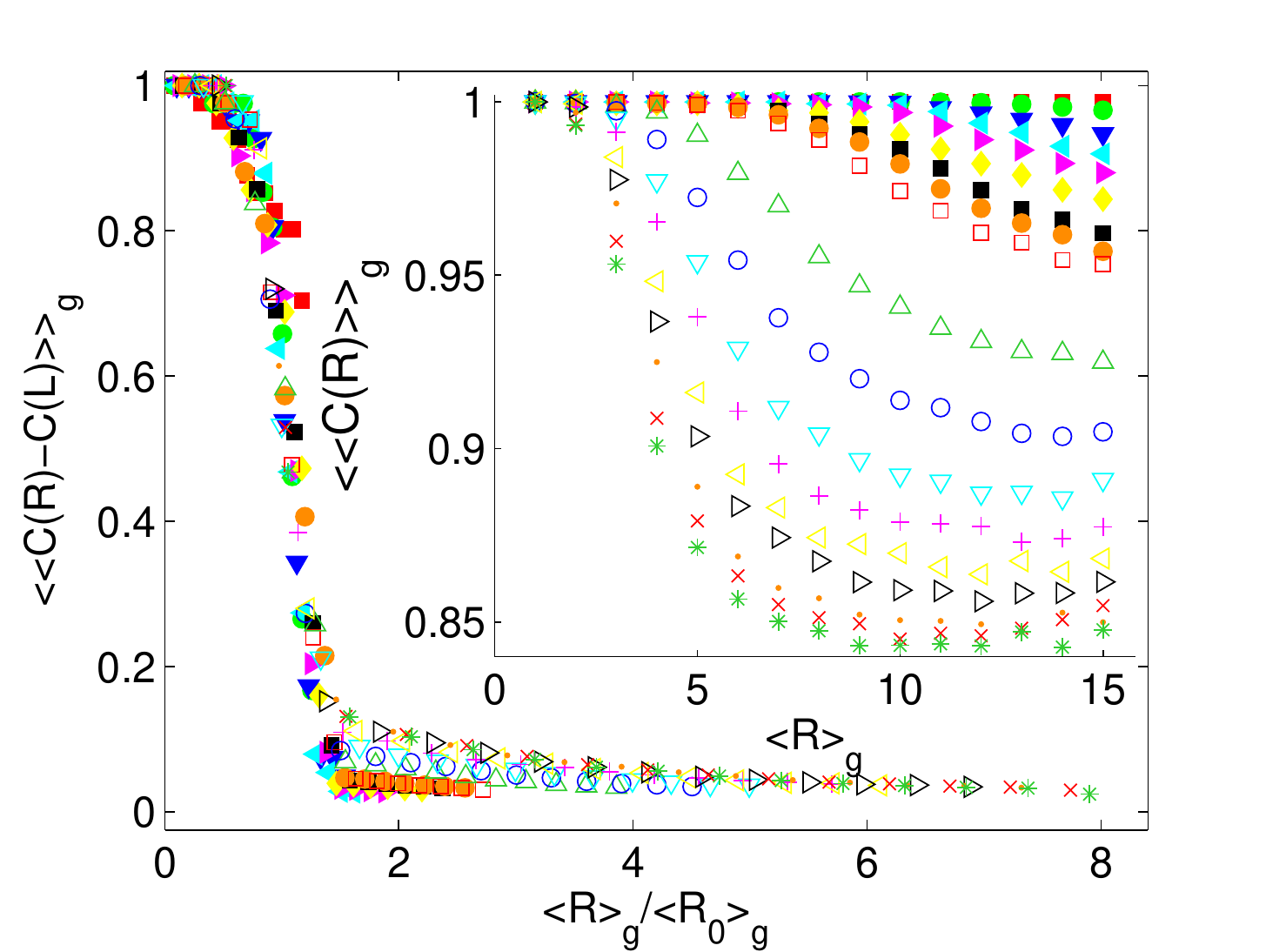}
\par\end{centering}

\caption{The connected correlation function for $M=900$. (Inset) The measured
PTS correlation function, with different colors and symbols corresponding
to different $\delta\phi$. Open symbols range from $0.01$ to $0.1$
(in increments of 0.01) and filled symbols range from $0.001$ to
$0.008$ (increments of 0.001). The highest $\delta\phi$, and hence
the highest pressures, correspond to the curves in the inset which
decay to the lowest values.\label{fig:C_rho_main_results}}
\end{figure}

In addition, $\left\langle \left\langle C\left(R\right)\right\rangle \right\rangle _{g}$
appears to asymptote to some value greater than zero for large $R$.
Subtracting off this asymptotic value $q_{0}=C\left(R=L/d\right)$,
for linear system size $L$, one can define a connected correlation
function $\left\langle \left\langle C\left(R\right)-q_{0}\right\rangle \right\rangle _{g}$.
When $R$ is scaled by $R_{0}$, the connected correlation function
collapses onto a master curve that decays rapidly to zero, a functional
form which RFOT theory predicts for the PTS correlation function \cite{BiroliAdamGibbs,Wolynes}. 

\subsection*{Deviation from the Mean-Field Exponent: Finite Size Scaling}

\begin{figure}
\subfigure[]{\includegraphics[scale=0.53]{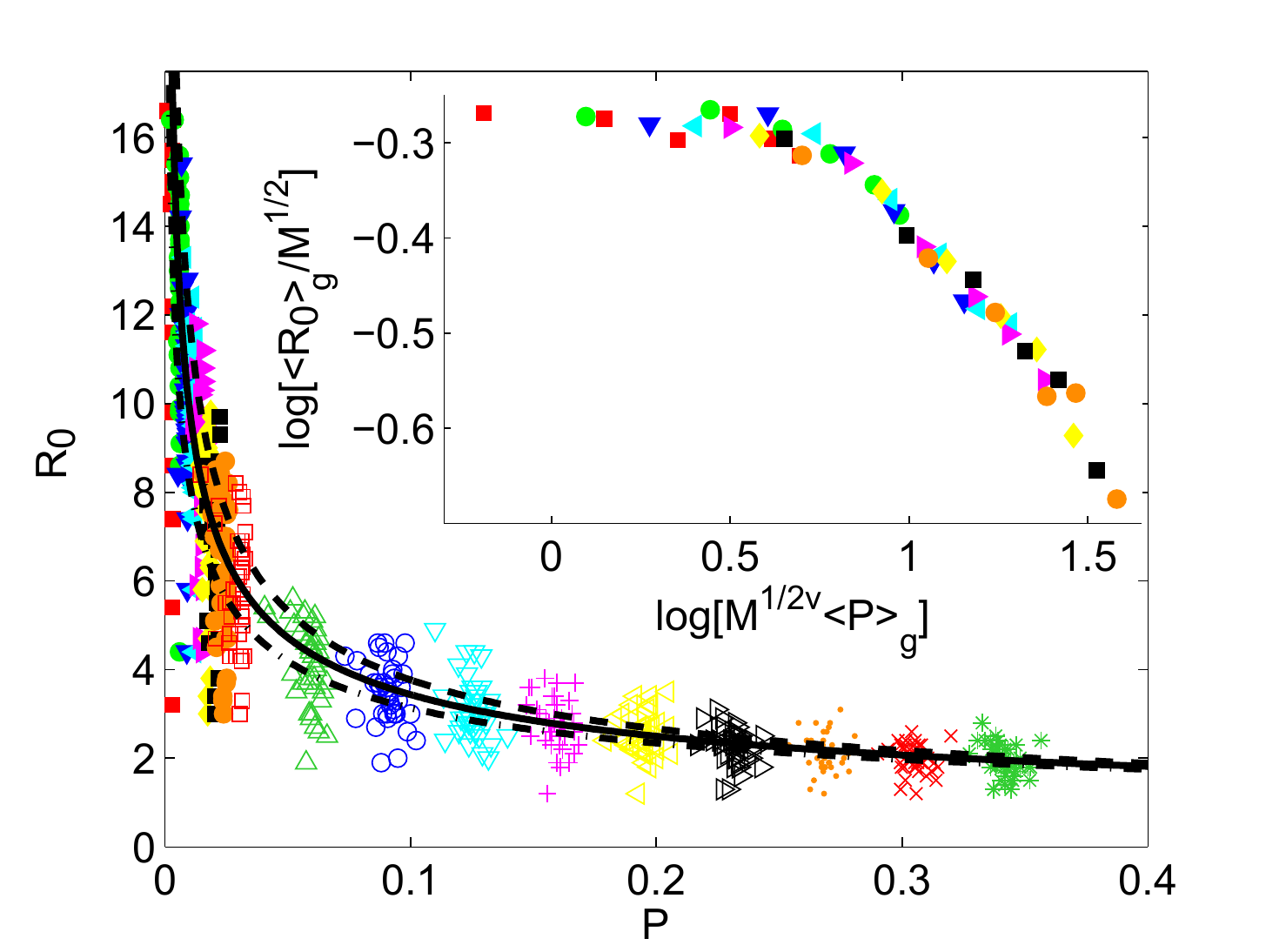}}\hfill{}\subfigure[]{\includegraphics[scale=0.53]{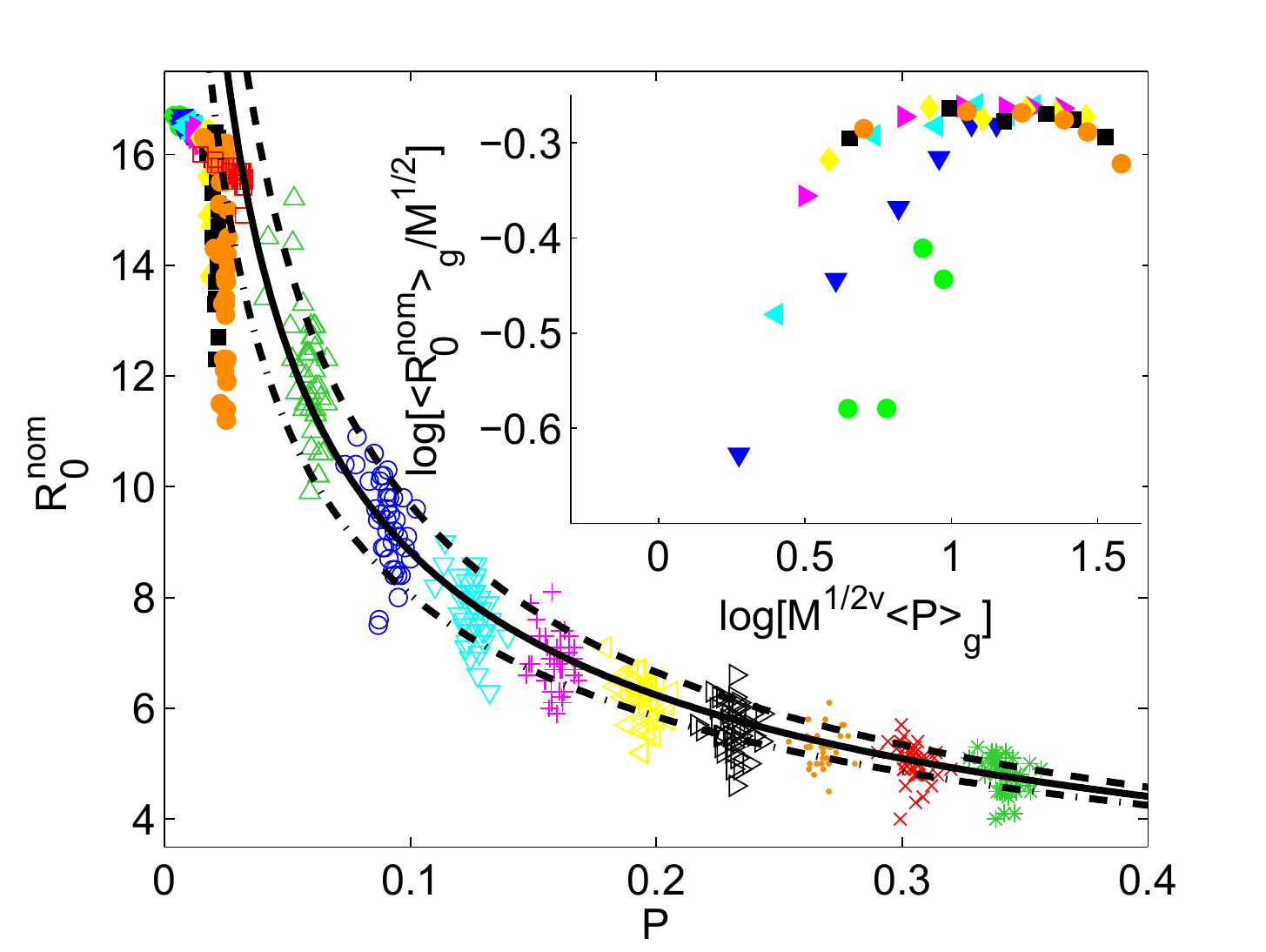}

}\caption{(a) The length scale $R_{0}$ is plotted versus $P$; a power law
fit is shown with a solid line, and dashed lines are used for $\nu=0.46 \pm0.04$. 
The inset shows the finite size scaling collapse for $\nu=0.46$.
(b) Same, but for the nominal length scale $R_{0}^{nom}$ and $\nu=0.5\pm0.04$, with the inset showing a collapse for $\nu=0.5$. Symbol colors are consistent with those used in figure \ref{fig:C_rho_main_results}.
\label{fig:power_law_plots_w_insets}}
\end{figure}

Since $\delta n\left(R<R_{0}\right)=0$, there is no null space
of $A\left(R\right)$.  The correlation function $C\left(R\right)$ for $R<R_{0}$ is identically 1 since $C\left(R<R_{0}\right)=\hat{f}_{0}\cdot\hat{f}_{0}=1$.
The length scale is extracted from the numerics by identifying that
value of $R$ at which $\delta n\left(R\right)$ is first greater
than zero (or the value of $R$ at which $C\left(R\right)$
is less than 1). As a function of $\left\langle P\right\rangle _{g}$,
the geometry-averaged critical length $\left\langle R_{0}\right\rangle _{g}$
fits a power law. A linear fit of $ln\left(\left\langle R_{0}\right\rangle _{g}\right)$
as function of $ln\left(\left\langle P\right\rangle _{g}\right)$
establishes an exponent of $\nu=0.461\pm0.012$ with a $95\%$ confidence
interval for the largest (900 grain) packings (figure \ref{fig:power_law_plots_w_insets}).
The lowest overcompression, $\delta\phi=0.001$, is left out of the
fit because nearly all of the geometries have $R_{0}=L$. The inclusion
of $\delta\phi=0.001$ changes the exponent to $\nu=0.45\pm0.015$.
Either way, while the exponent is near the mean-field value of $0.5$,
the difference is significant considering the confidence bounds. 

The exponent is verified using finite size scaling. Since all of the
numerical results are done at finite system sizes, the exponent that
is observed in power law fits may be susceptible to system size effects.
Furthermore, a length scale only truely diverges at a critical point
in the infinite system limit, so establishing a length scale which
becomes the system size in a finite system near the critical point
begs the question, is this length finite at the critical point but
just larger than any system size we have studied? The purpose of finite
size scaling is to isolate the functional form of the length scale
in the infinite system limit by scaling out the dependence on the
system size. First assume a form

\[
\left\langle R_{0}\right\rangle _{g}=L\cdot g\left(L^{1/\nu}\left\langle P\right\rangle _{g}\right)\]

\begin{figure}[t]
\subfigure[]{\includegraphics[scale=0.53]{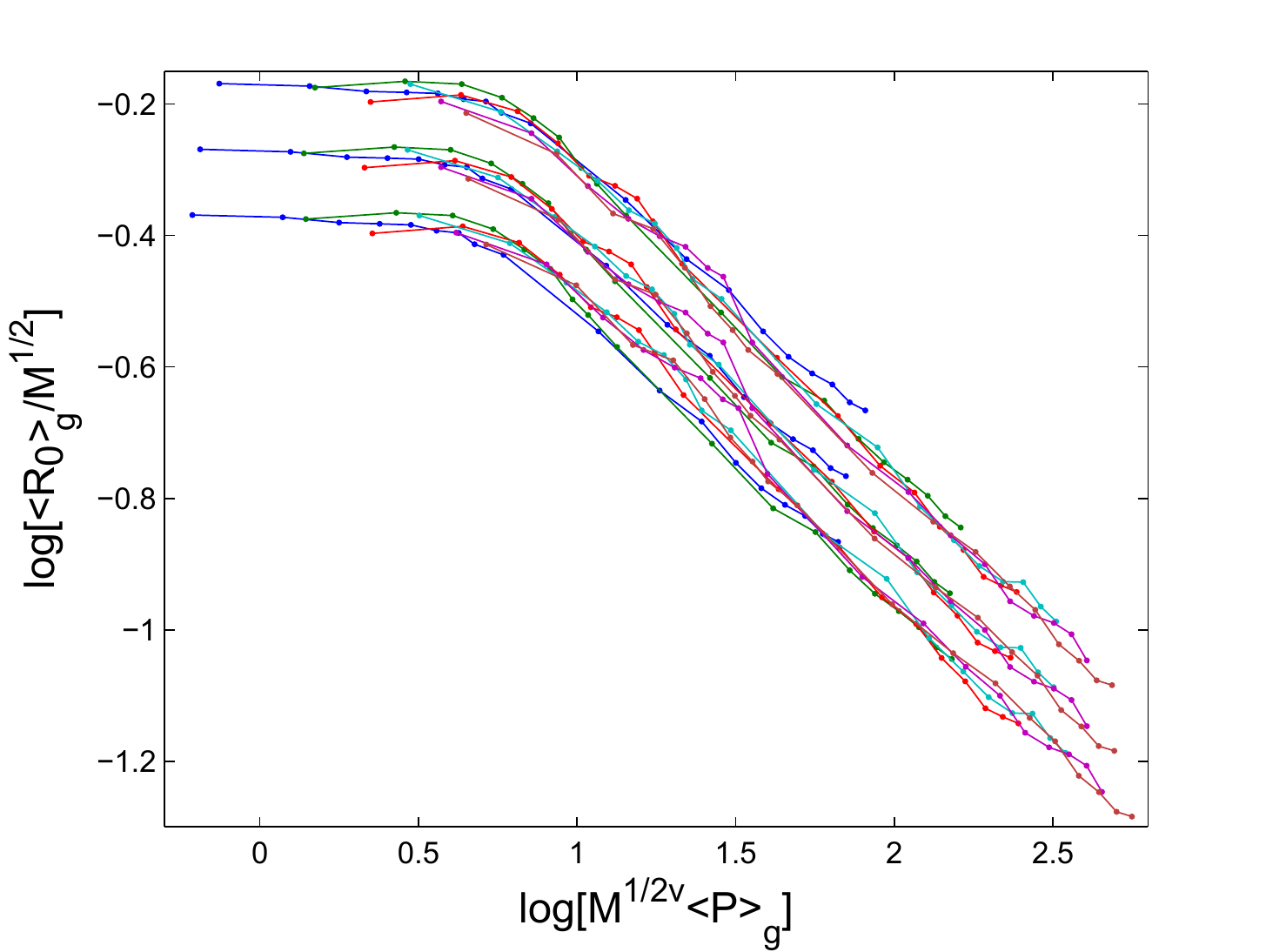}}\hfill{}\subfigure[]{\includegraphics[scale=0.53]{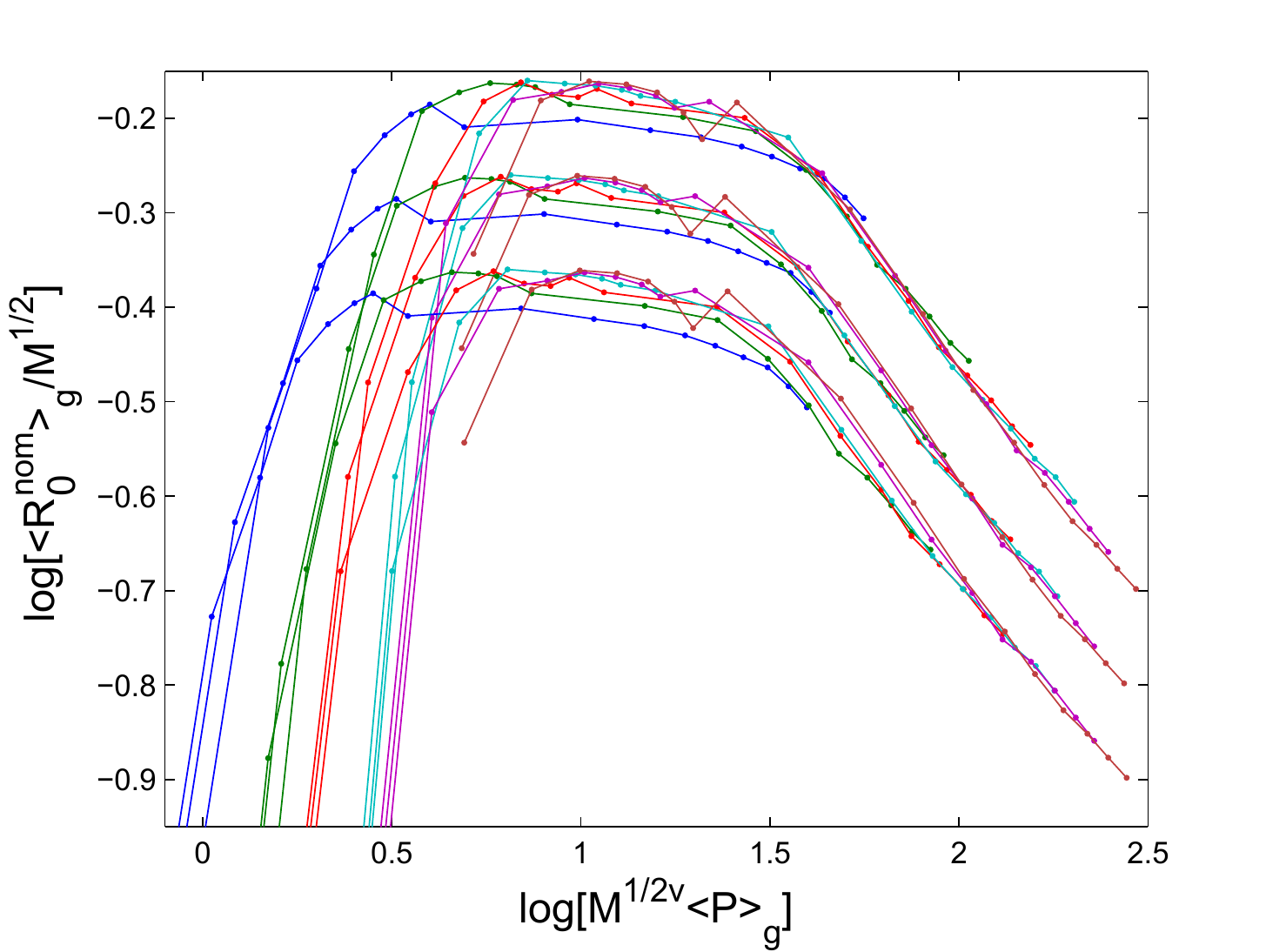}

}\caption{Presented in (a) is the finite size scaling collapse
for $\nu=0.46\pm0.04$. The lowest curve is for $\nu-0.04$, the middle
curve for $\nu\left(=0.46\right)$, and the top curve is for $\nu+0.04$.
The curves are artificially offset vertifcally and horizontally. Figure
(b) shows similar finite size scaling plots for $R_{0}^{nom}$,
with $\nu=0.50\pm0.04$. \label{fig:Exp_comp}}

\end{figure}

\noindent Here $L=\sqrt{M}$ has been used as a measure of the system
size. The scaling form should be the same for any system size, for an 
appropriate choice of the value of $\nu$. The above scaling has been studied
for several values of $\nu$ near the value given by the power law
fit, including the mean-field value. The inset of figure \ref{fig:power_law_plots_w_insets}
shows the finite size scaling collapse at low pressure. There is a
noticeable difference in the quality of the collapse for different
system sizes for $\nu=0.42,0.46$ and $0.50$ with the best collapse
being for $\nu=0.46$. Figure \ref{fig:Exp_comp} shows the finite
size collapse over the entire range of $P$ for these exponents. For
$\nu=0.42$, the collapse fails in the mid-range of the tail, while
for $\nu=0.50$ the collapse begins to fail in the knee of the scaling
form as well as the tail. The collapse for $\nu=0.46$ seems to be
the best comprimise between the two.  The deviation of the exponent from mean-field is significant but small, possibly reflecting logarithmic corrections, which we will explore further in section \ref{sec:Corrections-to-l*}.

To better understand the deviation from mean-field, let's return to
the bulk-surface argument discussed in \ref{sub:Bulk-Surface-Argument}.
The $R_{0}$ extracted from the numerics and used thus far in the
discussion of the exponent $\nu$ is found by identifying $\delta n\left(R_{0}\right)=0$,
where $\delta n$ is the nullity of the linear system that describes
the subregion of the packing of size $R$ found from the SVD. Let's
define a nominal nullity $\delta n_{nom}$ which is simply $n\left(R\right)-m\left(R\right)$,
where $n\left(R\right)$ is the number of contacts inside a boundary
of size $R$ and $m\left(R\right)$ is the number of grains
inside and on the boundary, so that $\delta n_{nom}$ is the expression for the mean field bulk-surface argument presented in Section \ref{sub:Bulk-Surface-Argument}.  If the mean field
approximation discussed in Section \ref{sub:Bulk-Surface-Argument} is valid,
the equality $\delta n=\delta n_{nom}$ should hold. In fact, $\delta n_{nom}$
does not equal $\delta n$, and what's more, a $R_{0}^{nom}$ extracted
from $\delta n_{nom}$ reproduces the mean-field result. Figures \ref{fig:power_law_plots_w_insets}
and \ref{fig:Exp_comp} also show the results for $R_{0}^{nom}$.
Since so much of the data at low pressures saturate at the system
size for $R_{0}^{nom}$, it is difficult to do a power law fit
as with $R_{0}$, even for the largest system size. The power laws
plotted for $R_{0}^{nom}$ are not taken from a fit. Instead, the
exponent $0.5$ is verified using finite size scaling only. For small
$M^{1/2\nu}\left\langle P\right\rangle _{g}$ none of the collapses
do well. If one focuses only on the plateau and tail, the collapse for
$\nu=0.46$ fails particularly in the plateau and near the knee, while
$\nu=0.54$ fails in the tail. $\nu=0.5$ seems to be the best balance
between the two. The finite size scaling for $R_{0}$ works well
over a larger range, including the plateau.

\begin{figure}%
\centering{}\includegraphics[scale=0.53]{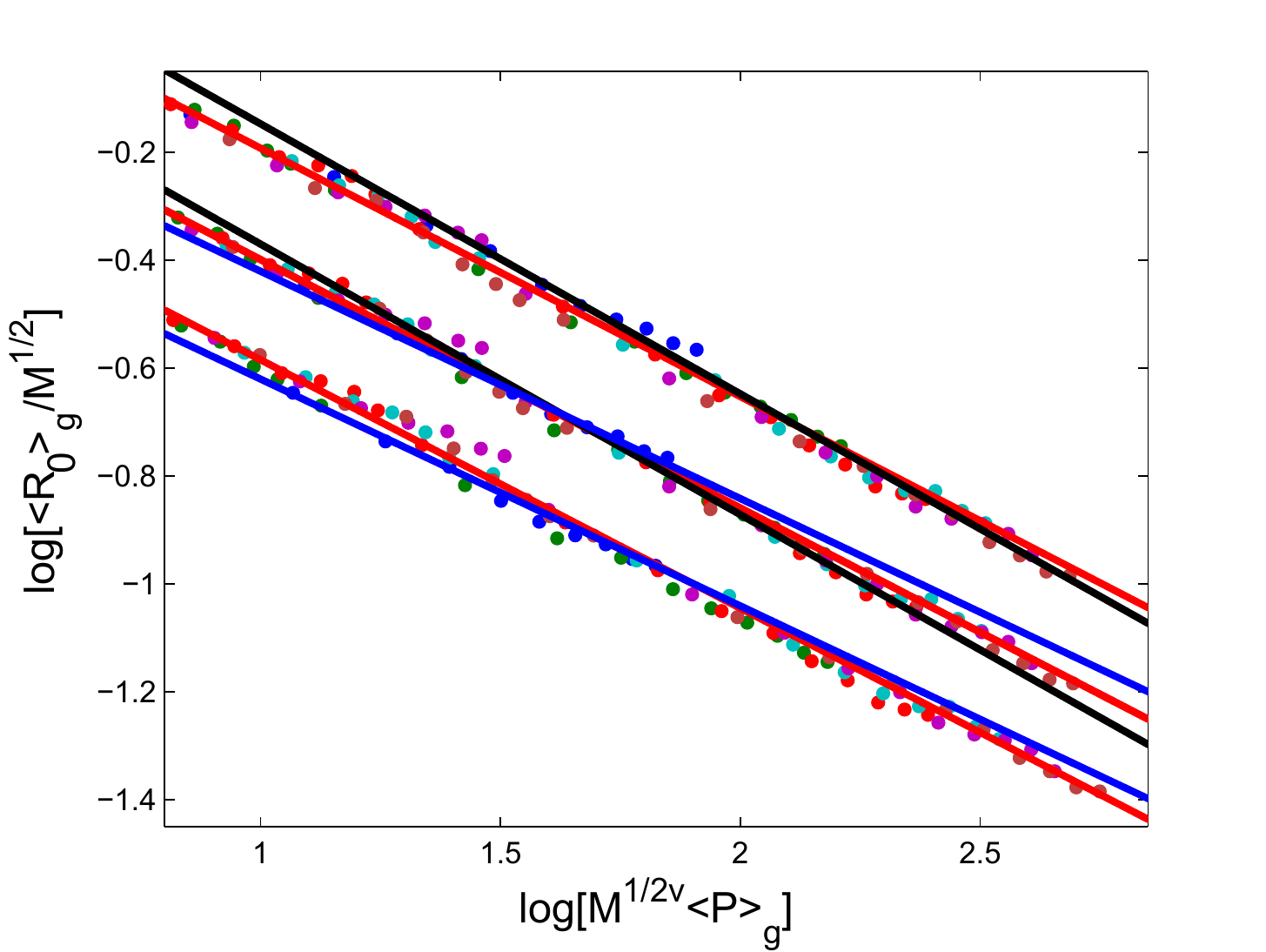}\caption{As with figure \ref{fig:Exp_comp}, the finite size scaling collapse
is shown. The lines connecting data points at the same system size
are no longer shown, but symbols use the same colors, and the same
vertical shift is used. Blue, red, and black lines have slopes 0.42,
0.46, and 0.50 respectively. \label{fig:Exp_compTails}}
\end{figure}

As an additional test of the the finite size scaling collapse used
to verify the exponent $\nu$, one can also focus on the tail of the
collapse. In the tail, one expects $\left\langle R_{0}\right\rangle _{g}/M^{1/2}$
to exhibit a power law dependence on $M^{1/2\nu}\left\langle P\right\rangle _{g}$.
If the exponent of this power law and the exponent used in the finite
size scaling differ, this leads to a failure of the collapse at different
system sizes. However, within the limits set by the variance in the
data due to the sensitivity of the collapse on the exponent used,
the tail of $\left\langle R_{0}\right\rangle _{g}/M^{1/2}$ should
have a slope corresponding to the infinite system size exponent. This
can be seen by assuming a power-law form for the scaling function,
and using $\nu_{t}$ as the {}``test'' exponent used to collapse
the finite system size data:

\[
\left\langle R_{0}\right\rangle _{g}=M^{\frac{1-\nu/\nu_{t}}{2}}\left\langle P\right\rangle _{g}^{-\nu}\]

\noindent Because $\nu_{t}$ is only applied to the system size variable
$M$, the pressure depends on $\left\langle R_{0}\right\rangle _{g}$
only through $\nu$. Figure \ref{fig:Exp_compTails} shows the finite
size collapse for the test exponent equal to 0.42, 0.46, and 0.50.
In all three cases, a line with a slope of $0.46$ seems to agree
best with the data, suggesting that $\nu=0.46$ is in indeed the exponent
in the infinite system size limit. 

\begin{figure}[t]
\subfigure[]{\includegraphics[scale=0.53]{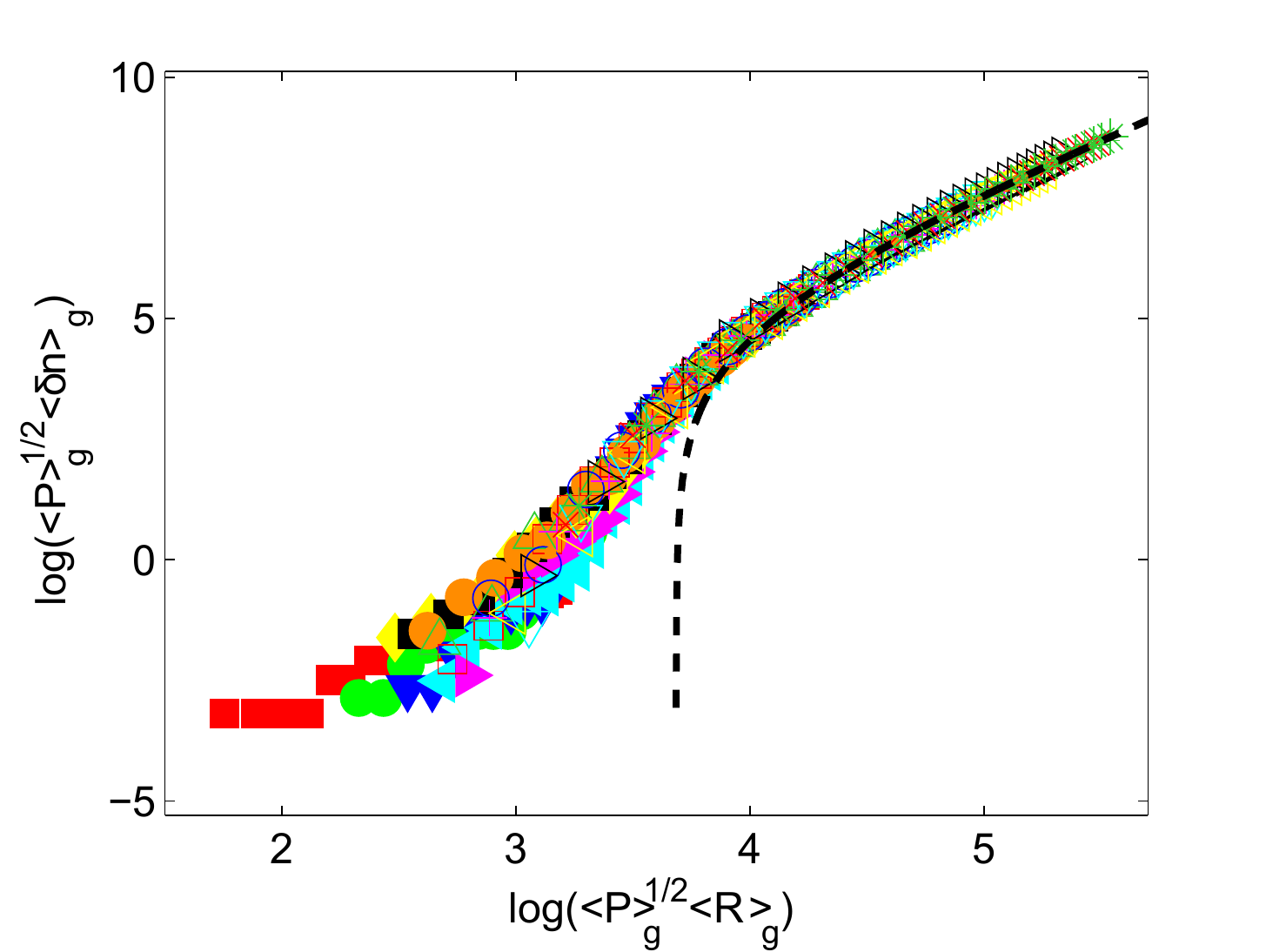}}\hfill{}\subfigure[]{\includegraphics[scale=0.53]{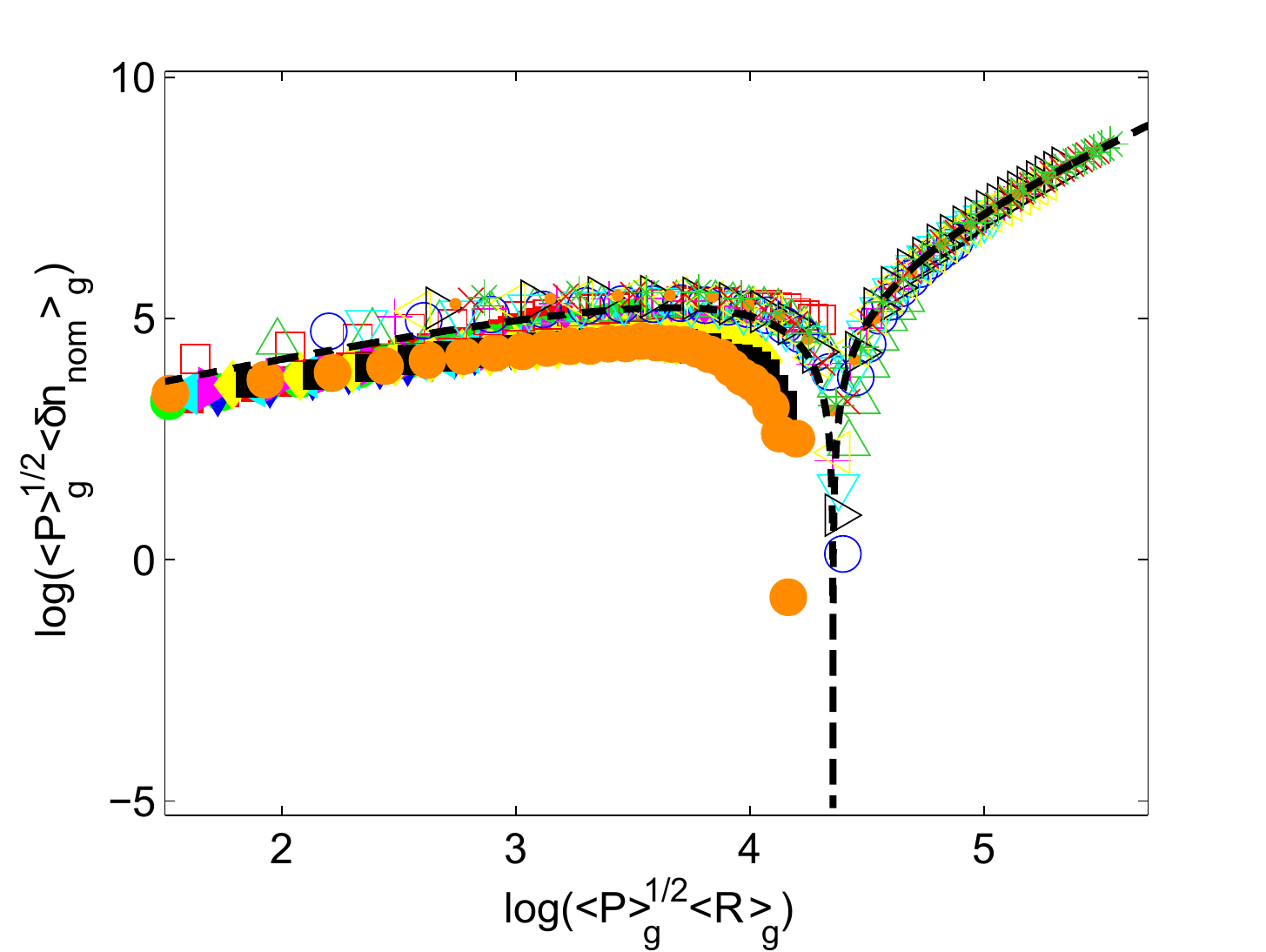}

}

\caption{(a) Collapse of $\delta n\left(R\right)$ after scaling with $P$.
The quadratic fit (dashed line) is not shown over the entire data
range because for low $P^{1/2}R_{0}$ the fit becomes negative,
while $\delta n\left(R\right)$ is always positive. (b) Same scaling
for $R_{0}^{nom}$. A quadratic fit agrees with the data well over
the entire data range.  Symbol colors are consistent with those used in figure \ref{fig:C_rho_main_results}.  As before, open symbols range from 0.01 to 0.1 and filled symbols from 0.001 to 0.008.\label{fig:dn_data}}
\end{figure}

The recovery of the mean-field exponent for $R_{0}^{nom}$ implies
something very interesting about the bulk-surface argument. The mean-field
result for $\delta n_{nom}$ as presented in Section \ref{sub:Bulk-Surface-Argument} has a form $\alpha P^{1/2}R^{2}-\beta R$.
Even if the nullity from SVD, $\delta n$, differed from $\delta n_{nom}$
by a term linear in $R$, the functional form of the length scale
wouldn't change. The difference in exponents for $R_{0}$ and $R_{0}^{nom}$
must be the result of a novel $P$ dependent term in the bulk-surface
argument for $\delta n$: $\delta n - \delta n_{nom}=P^{-1/2}f\left(P^{1/2}\right)$ for some unknown function f.  Figure \ref{fig:dn_data} suggests that
scaling forms for both $P^{1/2}\delta n$ and $P^{1/2}\delta n_{nom}$
exist and are functions of $\left\langle R\right\rangle _{g}\left\langle P^{1/2}\right\rangle _{g}$
only. But, the scaling form of $\delta n_{nom}$ is a quadratic over
the entire data range, even below the cusp at $\left\langle R_{0}^{nom}\right\rangle _{g}\left\langle P^{1/2}\right\rangle _{g}$,
while $\delta n$ fails to be quadratic near $\left\langle R_{0}\right\rangle _{g}\left\langle P^{1/2}\right\rangle _{g}$.
Section \ref{sec:Corrections-to-l*} discusses four distinct sources
of disagreement between $\delta n$ and $\delta n_{nom}$. 

\section{Modelling the PTS}

\subsection*{Geometry of the FNE Solution Space}

The set of solutions to Eq.\ref{eq:LinearSystem} sampled by the
random walk make up a high dimensional vector space. Two separate
but related spaces are discussed here. The more commonly discussed
space is the {}``force space,'' a $z$ dimensional space where each
axis corresponds to a different contact force magnitude\cite{FNE}.
Not every point in this space is a solution. However, there are bounds
on where solutions can be located in the force space. The positivity
constraint confines all solutions to the $\left\{ f_{1}...f_{z}\right\} \geq0$
hyper-octant. Additionally, a linear global constraint, say from fixing
the pressure, is generally applied to the FNE, so that there is at
least one constraint of the form $\sum_{i=1}^{z}c_{i}f_{i}=C$, where
the $c_{i}$'s and $C$ are some constants from the geometry of the
packing, for instance the center-to-center separations of the disks
for the pressure constraint. Ignoring the ME constraints for a moment,
the linear global constraints require that at least one $f_{i}$ be
non-zero. In principle, there can be solutions where $f_{i}=C$ while
$f_{j\neq i}=0$. This provides the point of intercept with each axis
of the force space. Since the global constraints are linear, each
intercept is connected by a line, forming a high dimensional polygon
(polytope). 

Contained within this polytope are all of the valid force networks
for the given MS, although much of the space within this polytope
does not correspond to valid solutions. The vertices, for instance,
which are the intercepts with the $f_{i}$ axes, cannot be solutions.
They correspond to force networks where $f_{i}=C>0$ and $f_{j\neq i}=0$,
which is never a force network that satisfies ME. 

It is possible though to apply the ME constraints by further changing
the shape and dimensionality of the polytope. Beginning with the polytope
defined by only the global constraint and some valid force network,
choose a ME constraint which is applied to grain $i$ with $z_{i}$
contacts. Such a constraint relates $z_{i}$ contact forces linearly,
so it is represented geometrically as a $z_{i}$ dimensional (unbounded)
polytope. Each ME constraint allows $f_{i}=0$ as a solution, 
so the
ME polytopes intersect the origin. Therefore, an ME polytope is never
entirely embedded in the global constraint polytope. Once imbedded
in the force space, the intersection of ${\it all}$ such ME constraint
polytopes with each other as well as the global constraint polytope
represents the set of valid force networks. The polytope that results
from these intersections contains what will be referred to as the
{}``solution space,'' equal in dimension to the nullity of $A$.
The solution space is simply connected because of the linearity of
the forces: the sum of any two force networks is also a valid force
network, and properly normalized will satisfy the global constraint. The solution space is also likely convex; \cite{FNE} claims that
the force space is {}``trivially'' convex, also due to the linearity,
but fails to take into account the positivity constraint when coming
to this conclusion.  The positivity constraints have similar implications for the solution space: while linear superpositions of valid force networks do result in valid force networks, they do not necessarily result in contact forces which are all positive.  However, an assumption of convexity allows us to estimate the number of valid force networks at a given pressure, as which we now explain in more detail.

Particular properties of the solution space are dependent on the MS.
But, a useful approximation is that the solution space is enclosed
by a hypersphere of dimension $\delta n$. Since, in FNE,  each valid force
network is considered to be equally likely, and the solution space
is simply connected, the volume of the $\delta n$-dimensional hypersphere
estimates the number of solutions. The radius of the hypersphere for solutions within a bounded region of linear size $R$
is $\left\langle \eta\left(R\right)\right\rangle $, the average
distance from $\vec{f}_{0}\left(R\right)$ to a point on the hypersphere.
When $\delta n>2$, a random walk through the solution space beginning
at $\vec{f}_{0}\left(R\right)$ is transient; that is to say, the
walker will rapidly approach the surface of the hypersphere, and spend
the majority of its time there. This reflects the geometric property
of high dimensional polytopes which is that the majority of the its
volume is concentrated near the boundary. As the dimension $\delta n$
becomes larger, a good approximation of $\left\langle \eta\left(R\right)\right\rangle $
results from sampling $\vec{f}\left(R\right)-\vec{f}_{0}\left(R\right)$
over the random walk steps: $\left\langle \eta\left(R\right)\right\rangle =\left\langle \left|\vec{f}\left(R\right)-\vec{f}_{0}\left(R\right)\right|\right\rangle $.
This is essentially a radius of gyration for the high-dimensional
random walk, and a similar quantity has been used previously as a
measure of force indeterminacy\cite{McNamara}. 

The volume of the hypersphere plays an essential role in the understanding
of the unjamming transition as a critical point. As $R \rightarrow R_0$, the hypersphere shrinks, both in linear size $\left\langle \eta\left(R\right)\right\rangle $
and in dimension, to a single point that is the only solution to the
precisely determined linear system of ME equations.  Since, as shown earlier, $R_0$ becomes the system size as $P\rightarrow0$,
if the volume of the hypersphere is a measure of the number of valid
force networks available, then the entropy of valid force networks
goes to zero as $P\rightarrow0$. The unjamming transition is associated
with a point at which there are no solutions available to Eq.\ref{eq:LinearSystem}.

\subsection*{Some Results Based on the Solution Space}

In this section we'll discuss a model of $\eta\left(R\right)$
and $C\left(R\right)$ based on properties of the solution space.
These results will be based on two approximations, concerning the
typical values for the {}``magnitudes'' of the force networks:

\begin{equation}
\mbox{\ensuremath{\left|\vec{\mathit{f}}_{\mathit{0}}\left(R\right)\right|}}=\left(\sum_{i=1}^{n\left(R\right)}f_{0,i}^{2}\right)^{1/2}\approx
n\left(R\right)^{1/2}\left[f_{0}\right]=\left|\vec{f}\left(R\right)\right|\label{eq:approx_mag}\end{equation}

\noindent and using $\vec{f}\left(R\right)=\vec{f}_{0}\left(R\right)+\sum_{i=1}^{\delta n\left(R\right)}c_{i}\hat{g}_{i}$

\begin{equation}
\eta\left(R\right)=\left|\vec{f}\left(R\right)-\vec{f}_{0}\left(R\right)\right|=\left|\sum_{i=1}^{\delta n\left(R\right)}c_{i}\hat{g}_{i}\right|=\left(\sum_{i=1}^{\delta
n\left(R\right)}c_{i}^{2}\right)^{1/2}\approx\sqrt{\delta n\left(R\right)}\left[f\right]\label{eq:eta_model}\end{equation}

\noindent The $f_{i}$'s are individual contact forces in the initial
force network $\vec{f}_{0}$. The brackets $\left[\ \right]$ represent
an average over elements of force networks
(or geometry). For instance, $\left[f\right]$ is the average contact
force, and $\left[z\right]$ is the average contact number for a particular
MS. There are several assumptions used here. First, the contact forces
themselves are uncorrelated; the sum over $n\left(R\right)$ contact
forces is simply $n\left(R\right)$ times the average contact force.
This assumption is particularly good for large force networks, and
is based on the observation that contact forces do not seem to exhibit
correlations beyond the size of a grain, where of course ME constraints
limit what values they can take (\cite{SilkeLongPaper} provides some
arguments that support this observation. It is shown that the assumption
of a uniform sampling of states leads to the decorrelation of contact
forces at the isostatic point). The second assumption that is made
is that all force networks at a given $P$ have the same magnitude:
$\mbox{\ensuremath{\left|\vec{\mathit{f}}_{\mathit{0}}\left(R\right)\right|}}=\left|\vec{f}\left(R\right)\right|$.
This assumption is based on the fact that the the pressure is fixed,
so that the sum of the contact forces is fixed. Of course this is
not equivalent to saying that the magnitudes (sum of squares of the
contact forces) cannot fluctuate, but those fluctuations should still
be controlled by $P$. Finally, the assumption is made that the coefficients
$c_{i}$ of the expansion of $\vec{f}\left(R\right)-\vec{f}_{0}\left(R\right)$
in terms of the null space basis vectors are also uncorrelated and
on average are equal to the average contact force, since $\left[f\right]$
is the only force scale in the system. 

The correlation function $C\left(R\right)$ is related to $\eta\left(R\right)$,
which can be seen by expanding $\eta\left(R\right)^{2}=\left|\vec{f}\left(R\right)-\vec{f}_{0}\left(R\right)\right|^{2}$:

\begin{equation}
\left|\vec{f}\left(R\right)-\vec{f}_{0}\left(R\right)\right|^{2}=f\left(R\right)^{2}+f_{0}^{2}\left(R\right)-2\vec{f}\left(R\right)\cdot\vec{f}_{0}\left(R\right)
\end{equation}

\noindent so that

\begin{equation}
C\left(R\right)\approx1-\frac{\eta\left(R\right)^{2}}{2f\left(R\right)^{2}}\label{eq:relate_eta_to_C}
\end{equation}

\noindent Using the results Equations \ref{eq:approx_mag} and \ref{eq:eta_model},
the correlation function $C\left(R\right)$ is 

\begin{equation}
C\left(R\right)\approx\frac{1}{2}\left(1+\frac{m\left(R\right)}{n\left(R\right)}\right)\label{eq:model_C}
\end{equation}

\noindent This is a valuable result because it suggests that the
correlation function can be evaluated by simple counting: $m\left(R\right)$
is the number of independent equations, and $n\left(R\right)$
the number of independent degrees of freedom, of $A\left(R\right)\vec{f}\left(R\right)=\vec{b}\left(R\right)$.
In the mean-field approximations of section \ref{sub:Bulk-Surface-Argument},
$m\left(R\right)$ should be (twice) the number of grains in the
subregion of size $R$, and $n\left(R\right)$ should be the
number of contact forces. 

The approximate form of the correlation function also explains a peculiar
property of the measurements of $\left\langle \left\langle C\left(R\right)\right\rangle \right\rangle _{g}$
from section \ref{sub:The-Correlation-Function}: the correlation
function does not decay to zero. In the $R\rightarrow\infty$ limit,
and in mean-field, $C\left(R\right)=\frac{1}{2}\left(1+\frac{2M}{z}\right)$,
which is $1$ at the isostatic point but is never less than $1/2$.
This is essentially the result of the positivity constraint. If contact
forces could be negative, $\left[f\right]$ would be zero when contact
forces were uncorrelated. The average must always be greater than
zero for compressive forces. Comparison of Eq. \ref{eq:model_C} with
numerical results are presented in section \ref{sec:Discussion}.

\section{Corrections to $l^{*}$ \label{sec:Corrections-to-l*}}

\subsection*{Logarithmic Corrections}

\begin{figure}[t]
\begin{centering}
\includegraphics[scale=0.53]{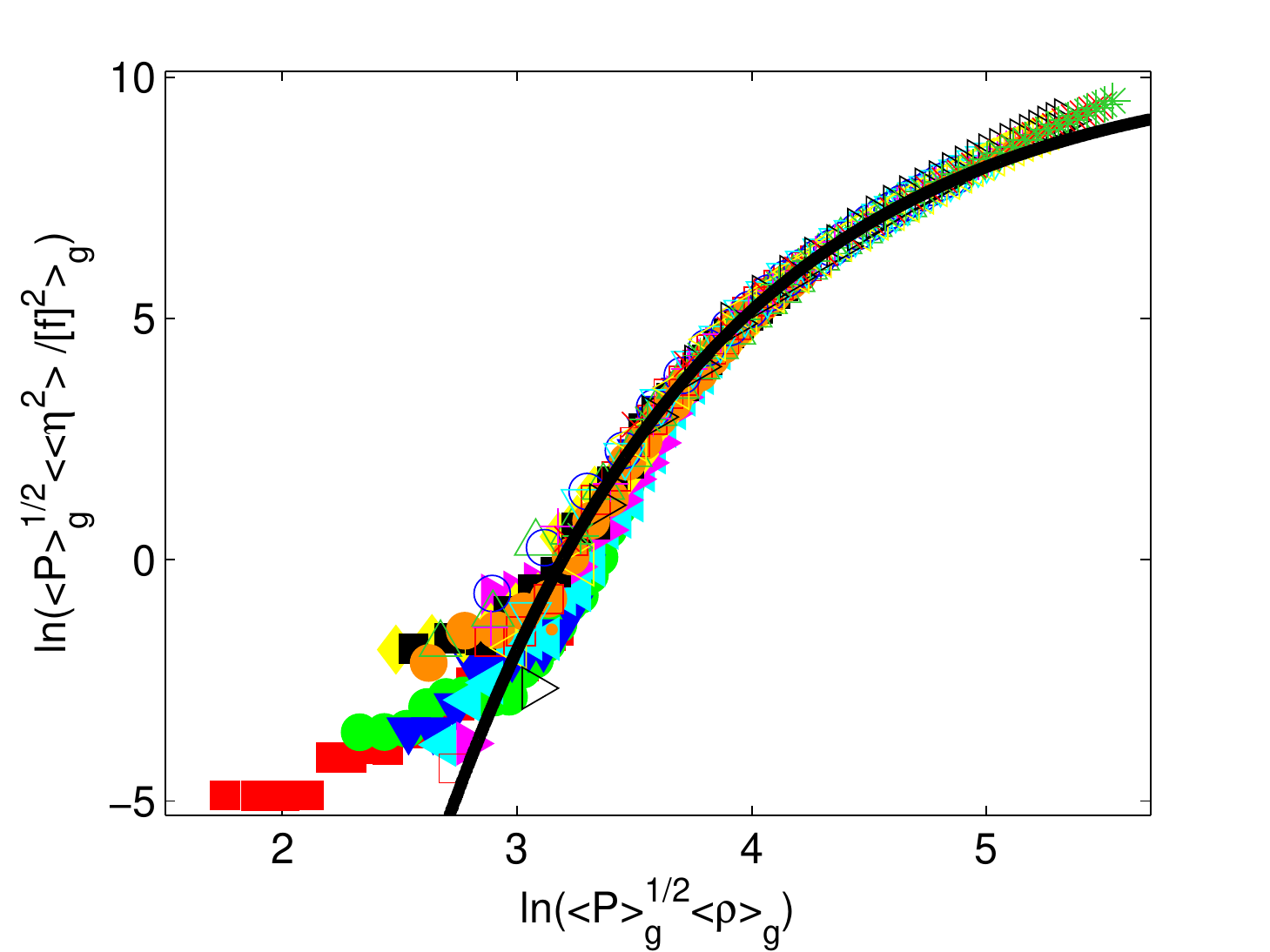}
\par\end{centering}

\caption{The collapse of $\bar{\eta}^{2}\left(R\right)$ with pressure is
shown. The solid line is a fit to $G\left(y\right)=G_{0}e^{-\left(b/y\right)^{\alpha}}$,
with $b=161$, $\alpha=0.86$, and $G_{0}=3\times10^{4}$. Notice
the similarity of this collapse to figure \ref{fig:dn_data}a, which
illustrates the equality of $\bar{\eta}$ and $\delta n$ (equation
\ref{eq:eta_model}). Symbol colors are consistent with those used in figure \ref{fig:C_rho_main_results}.\label{fig:scaling_of_eta_bar}}

\end{figure}

The deviation from the mean-field exponent of $\nu=0.5$ suggests
that mean-field bulk-surface argument breaks down. In this section
analysis of the behavior of $\eta\left(R\right)$ is used to identify
a logarithmic correction to the mean-field prediction. The radius
of gyration normalized by the force scale, $\bar{\eta}^{2}\left(R\right)=\eta^{2}\left(R\right)/\left[f\right]^{2}$,
collapses for different values of $P$ according to the scaling form
$\left\langle P\right\rangle _{g}^{1/2}\left\langle \bar{\eta}^{2}\left(R\right)\right\rangle _{g}=G\left(\left\langle R\right\rangle _{g}\left\langle P\right\rangle
_{g}^{1/2}\right)$
(see figure \ref{fig:scaling_of_eta_bar}). 

The hypersphere that encloses the valid force networks has a dimensionality
$\delta n\left(R\right)$, which has been shown (Eq.\ref{eq:eta_model})
to depend on $\bar{\eta}\left(R\right)$. Furthermore, the linear
size of the hypersphere is $\eta\left(R\right)$. In order for
the volume of the hypersphere to be a count of solutions to $A\left(R\right)\vec{f}\left(R\right)=\vec{b}\left(R\right)$,
the linear size must be unitless, and so the unitless volume $V\left(\left\langle \bar{\eta}\right\rangle _{g}\right)$
depends only on the average ${\it normalized}$ radius of gyration.
The number of solutions $V\left(\left\langle \bar{\eta}\right\rangle _{g}\right)$
approaches 1 as $\left\langle \eta\right\rangle $ decreases; the
hypersphere shrinks in size and in dimension. But, there is always
a single solution to ME, even at the isostatic point (where the linear
system is precisely determined). Therefore, the entropy $\ln\left(V\left(\left\langle \bar{\eta}\right\rangle _{g}\right)\right)$
goes to zero when $\left\langle \bar{\eta}\right\rangle _{g}$ goes
to some small value $\bar{\eta}_{0}$. 

\begin{figure}[t]
\begin{centering}
\includegraphics[scale=0.53]{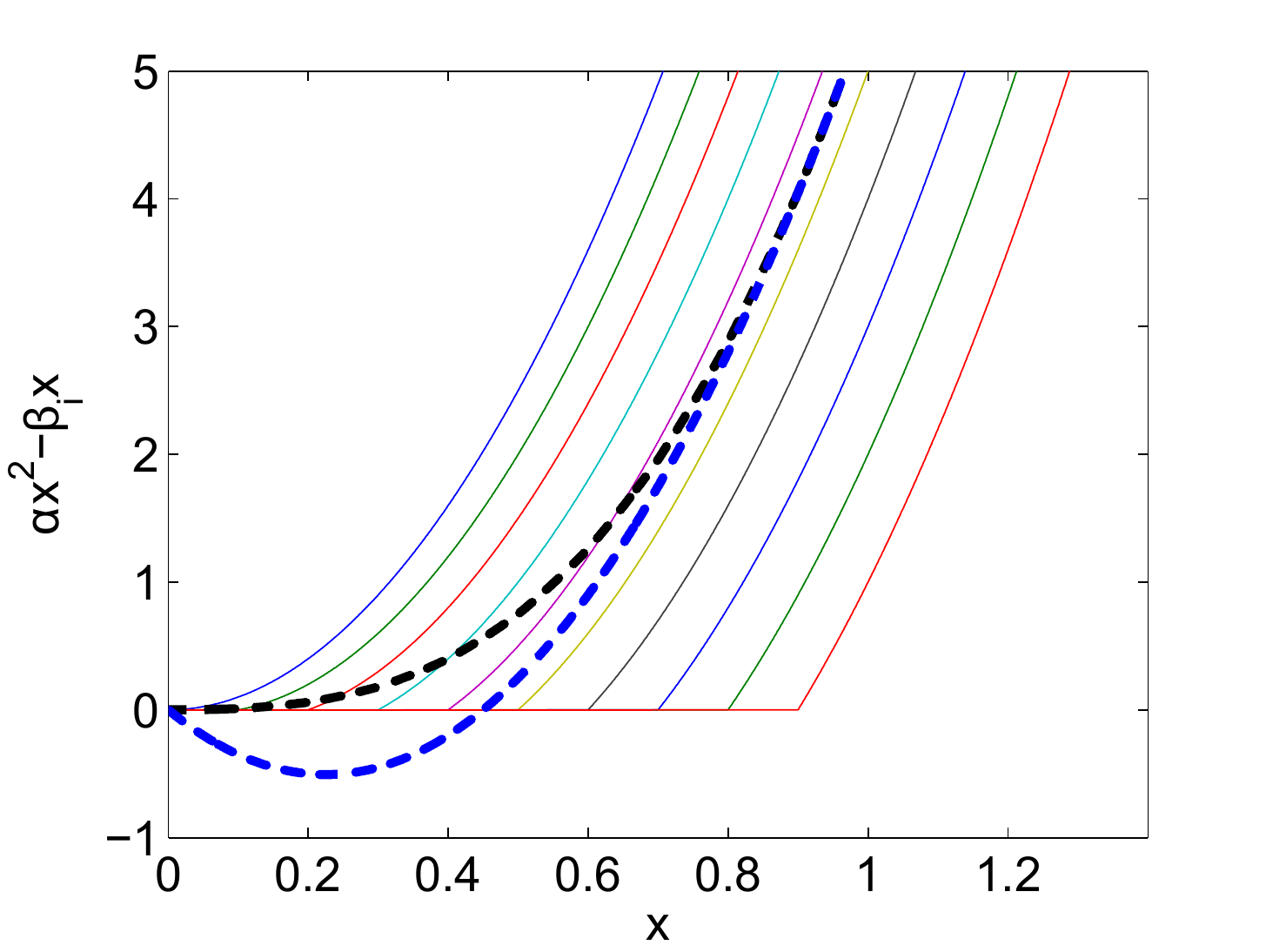}
\par\end{centering}

\caption{Ten truncated quadratic functions are shown. Below the root $\alpha/\beta_{i}$,
each quadratic function is set to zero. The average of the functions
is shown with a black dashed line, while the average of the untruncated
quadratics is shown with a blue dashed line. The horizontal intercept
of the black curve is at $x=0$, but at large $x$ the black curve
has a quadratic form. \label{fig:SchematicQuadAvg}}
\end{figure}

It should be mentioned that $\bar{\eta}_{0}$ only has an interpretation
as an MS-averaged value. For any particular MS, $\bar{\eta}$ is always
zero when $R<R_{0}$. But, over many geometries at a particular
$\left\langle P\right\rangle _{g}$, the value of $R_{0}$ can
fluctuate, and so $\left\langle \bar{\eta}\right\rangle _{g}$ is
only zero when $R=0$. On the other hand, the volume of the hypersphere
$V\left(\left\langle \bar{\eta}\right\rangle _{g}\right)$ becomes
1 at a value of $\left\langle R\right\rangle _{g}$ larger than
zero since it is a function of $\left\langle \bar{\eta}\right\rangle _{g}$
and not $\bar{\eta}$. This value of $\left\langle R\right\rangle _{g}$
corresponds to a value of $\left\langle \bar{\eta}\right\rangle _{g}$
which is greater than zero, but which controls the entropy-vanishing
behavior of the solution space, and it is this value of $\left\langle \bar{\eta}\right\rangle _{g}$
that is being referred to as $\bar{\eta}_{0}$ (see figure \ref{fig:SchematicQuadAvg}).

Next we look more closely at the scaling function $G\left(\left\langle R\right\rangle _{g}\left\langle P\right\rangle _{g}^{1/2}\right)$.
For small values of $y=\left\langle R\right\rangle _{g}\left\langle P\right\rangle _{g}^{1/2}$,
the scaling function is fitted well by $G\left(y\right)=G_{0}e^{-\left(b/y\right)^{\alpha}}$,
for fitting parameters $\alpha$, $G_{0}$, and $b$ (see caption for figure \ref{fig:scaling_of_eta_bar}). At $\left\langle \bar{\eta}\right\rangle _{g}=\bar{\eta}_{0}$,
the form of the scaling function implies 

\begin{equation}
\left\langle P\right\rangle _{g}^{1/2}\bar{\eta}_{0}^{2}=G_{0}exp\left\{ -\left(\frac{b}{\left\langle R_{0}\right\rangle _{g}\left\langle P\right\rangle
_{g}^{1/2}}\right)^{\alpha}\right\} \label{eq:scaling_of_eta_0}\end{equation}

\noindent and solving for the length scale $\left\langle R_{0}\right\rangle _{g}$, 

\begin{equation}
\left\langle R_{0}\right\rangle _{g}^{-1}=\frac{\left\langle P\right\rangle _{g}^{1/2}}{b}\left(ln\left\{ \frac{G_{0}}{\bar{\eta}_{0}^{2}}\right\} -\frac{ln\left\{ \left\langle
P\right\rangle _{g}\right\} }{2}\right)^{1/\alpha}\label{eq:log_corrections_to_length}\end{equation}

\noindent When $\left\langle P\right\rangle _{g}$ is large, which
in this case means relative to $G_{0}$, $\left\langle R_{0}\right\rangle _{g}\propto\left\langle P\right\rangle _{g}^{-1/2}$,
recovering the mean-field result. However, when $\left\langle P\right\rangle _{g}$
is small, the the length scale exhibits a logarithmic correction: 

\begin{equation}
\left\langle R_{0}\right\rangle _{g}\propto\left\langle P\right\rangle _{g}^{-1/2}\left(ln\left\{ \left\langle P\right\rangle _{g}^{-1/2}\right\}
\right)^{-1/\alpha}\label{eq:logcorrection}\end{equation}

\noindent This result is based on the observation that for small
$y$, the scaling function has an expontential form. This itself is
at odds with the mean-field prediction. The product $\left\langle P\right\rangle _{g}^{1/2}\left\langle \bar{\eta}^{2}\left(R\right)\right\rangle _{g}$
should be proportional to $\delta n\left(R\right)$, which mean-field
predicts should be a quadratic function of $R$, and so $G\left(y\right)$
should be quadratic in $y$ as well. For large $y$, the scaling function
is quadratic in $y$ and so the mean-field prediction $\left\langle R_{0}\right\rangle _{g}\propto\left\langle P\right\rangle _{g}^{-1/2}$
is recovered. For small $y$ on the other hand, the numerical results
show that the mean-field bulk-surface argument breaks down and $G\left(y\right)$
takes on an exponential form.  For very small $y$, the quality of the collapse breaks down and correspondingly the fit becomes less convincing, but the deviation from the quadratic form has already emerged (compare to figure \ref{fig:dn_data}a).

The following sections explore some microscopic origins for the failure
of the bulk-surface argument by analyzing the statistics of some microscopic
properties of the MS configurations, as well as some sample packing
geometries and the null spaces of their corresponding linear systems.

\subsection*{Local corrections to the Bulk-Surface Argument\label{sec:local_corrections_to_MF}}

\begin{figure}[t]
\subfigure[]{\includegraphics[scale=0.27]{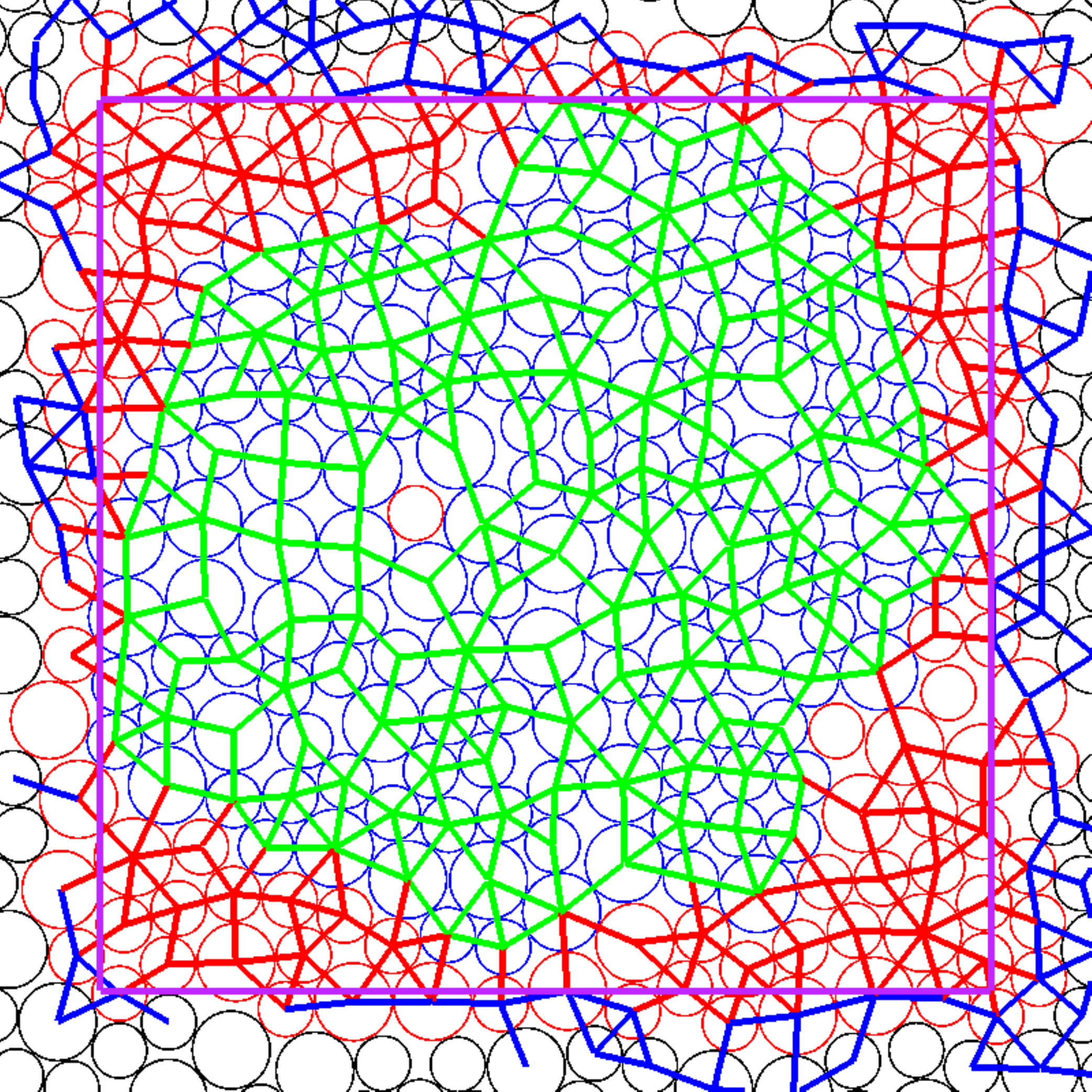}}\hfill{}\subfigure[]{\includegraphics[scale=0.27]{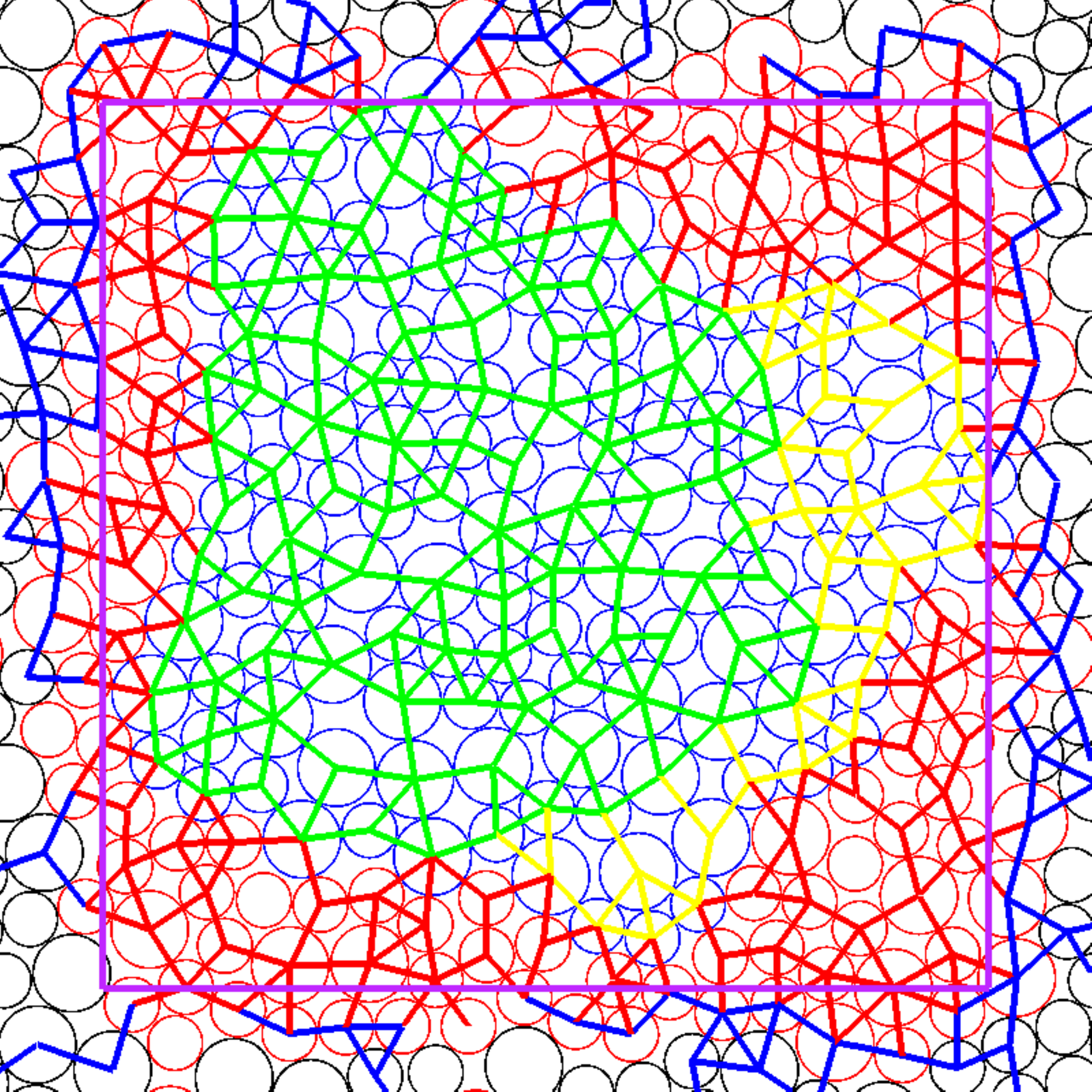}

}\caption{Figure a.) shows a 300 grain packing at $\delta\phi=0.01$. Locally
determined regions appear propagating inward from the boundary in
red. b.) The same packing at $\delta\phi=0.006$ shows the onset of
an isostatic cluster in yellow. \label{fig:Example_effective_constraints}}
\end{figure}

\begin{figure}[t]
\begin{centering}
\includegraphics[scale=0.65]{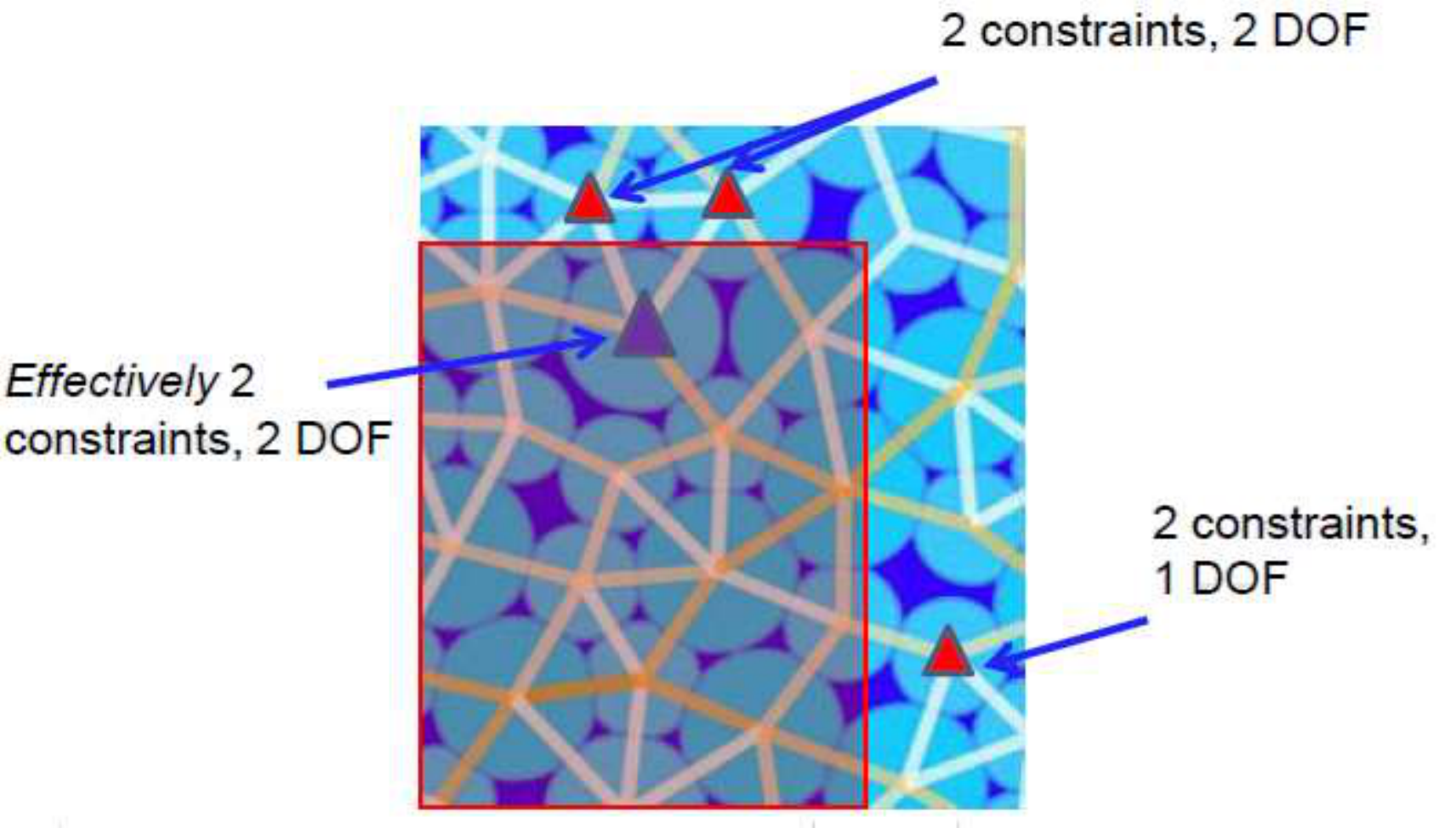}
\par\end{centering}

\caption{The red box in this example represents the boundary. While each grain
corresponds to two ME constraints, sometimes a grain corresponds to
less than three variable contact forces, such as the grains labelled
with red triangles. The grain with a purple triangle has 2 constraints
and ${\it effectively}$ only two variable contact forces even though
it has four contacts which lie inside the boundary. This is because
two of the contact forces are already determined by the additional
constraints from the boundary grains.\label{fig:Example_boundary_determined}}

\end{figure}

The first correction to the mean-field bulk-surface argument results
from the vectorial nature of the ME equations. Unlike most constraint
satisfaction problems, where each vertex or node contributes a single
constraint on the variables which correspond to links associated with
that vertex, with ME multiple equations describe force balance, further
coupling the force degrees of freedom. Specifically, in 2D, there
are two force balance equations associated with each grain. If a grain
has only one force exerted on it, that force variable is $\emph{overdetermined}$.
One equation is all that is necessary to solve for the force variable,
and the other equation is not necessary. This single force variable
is known and is effectively not a variable at all. In 2D, the same
happens when a grain has two force variables. The two equations of
ME can be used to define both force variables. Only grains with three
or more contacts are undetermined. In a packing without a fixed boundary,
there are never grains with less than three contacts. However, when
there is a boundary defined by a set of additional boundary constraints,
grains with only one or two force variables are possible (see figure
\ref{fig:Example_boundary_determined}). 

At low pressure the effect of the boundary grains which involve one
or two force variables also tend to completely define the contact
forces of neighboring grains. This allows for the phenomenon of effectively
defining force variables near the boundary to propagate into the interior
of the subregion, and at lower pressures this effect propagates further
into the interior (see figure \ref{fig:Example_effective_constraints}).
A recursive algorithm is used to find these effectively determined
force variables. All of the grains with two or less force variables
are found, and then all remaining grains, with two or less forces after
taking into account these effectively determined forces, are found.
This process is repeated until all effectively determined force variables
are found. 

\begin{figure}[t]
\centering{}\includegraphics[scale=0.55]{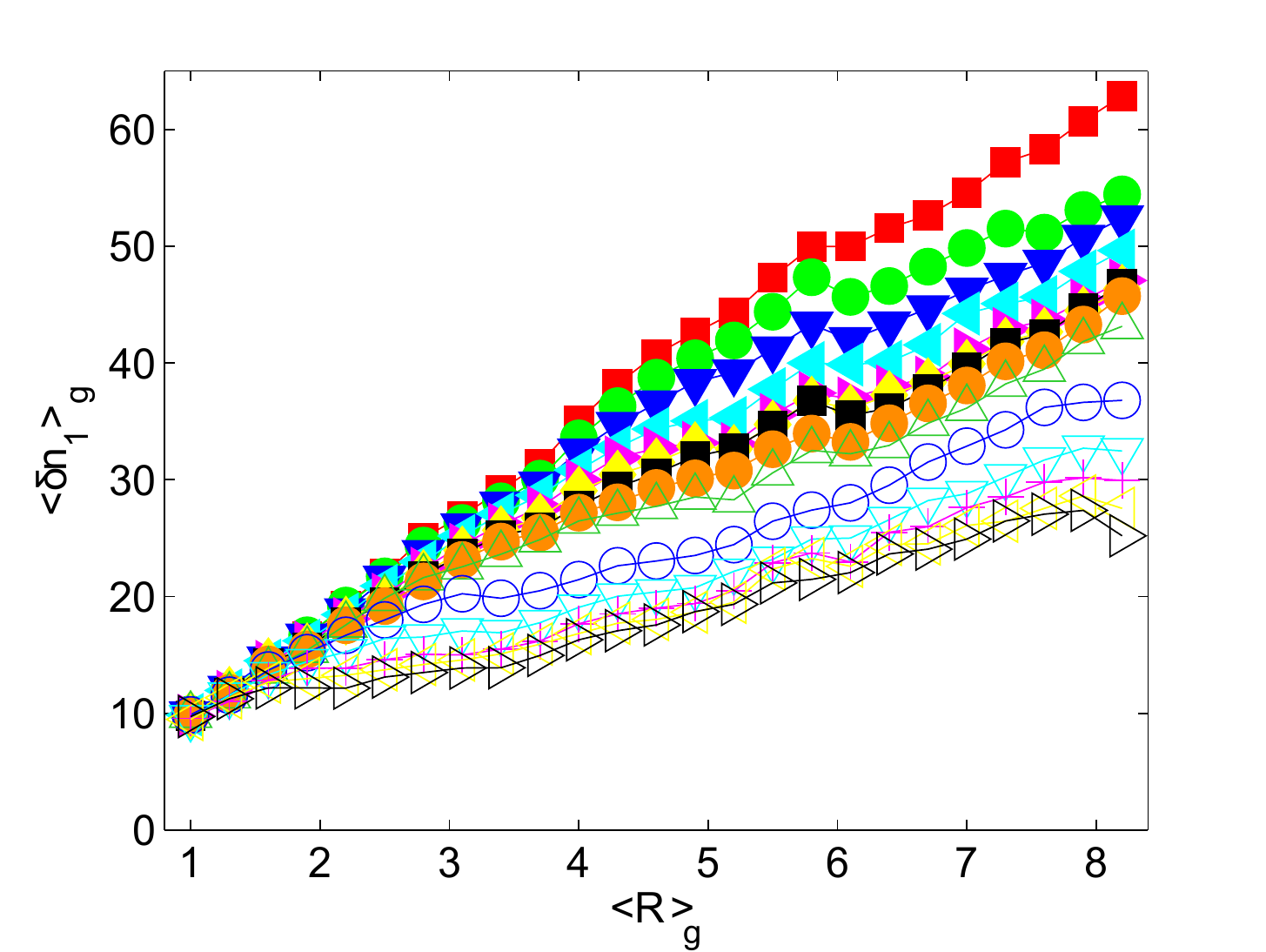}\caption{The local correction, $\delta n_{1}$, is the difference between the
number of ME equations and the number of determined force variables
(the red regions of the packings). The curves are not quite linear,
and depend on $P$ in a non-trivial way. There doesn't appear to be
a single scaling, although different ranges of data can be scaled
with $P$.Symbol colors are consistent with those used in figure \ref{fig:C_rho_main_results}. \label{fig:The-local-correction_stats}}
\end{figure}

These effectively determined forces and corresponding ME equations
both contribute corrections to the bulk-surface argument. The dependence
of these corrections is roughly linear as a function of $R$, but
the slope has a dependence on $P$ (figure \ref{fig:The-local-correction_stats}). 

\subsection*{Cooperative corrections\label{sec:Cooperative-corrections}}

\begin{figure}
\subfigure[]{\includegraphics[scale=0.27]{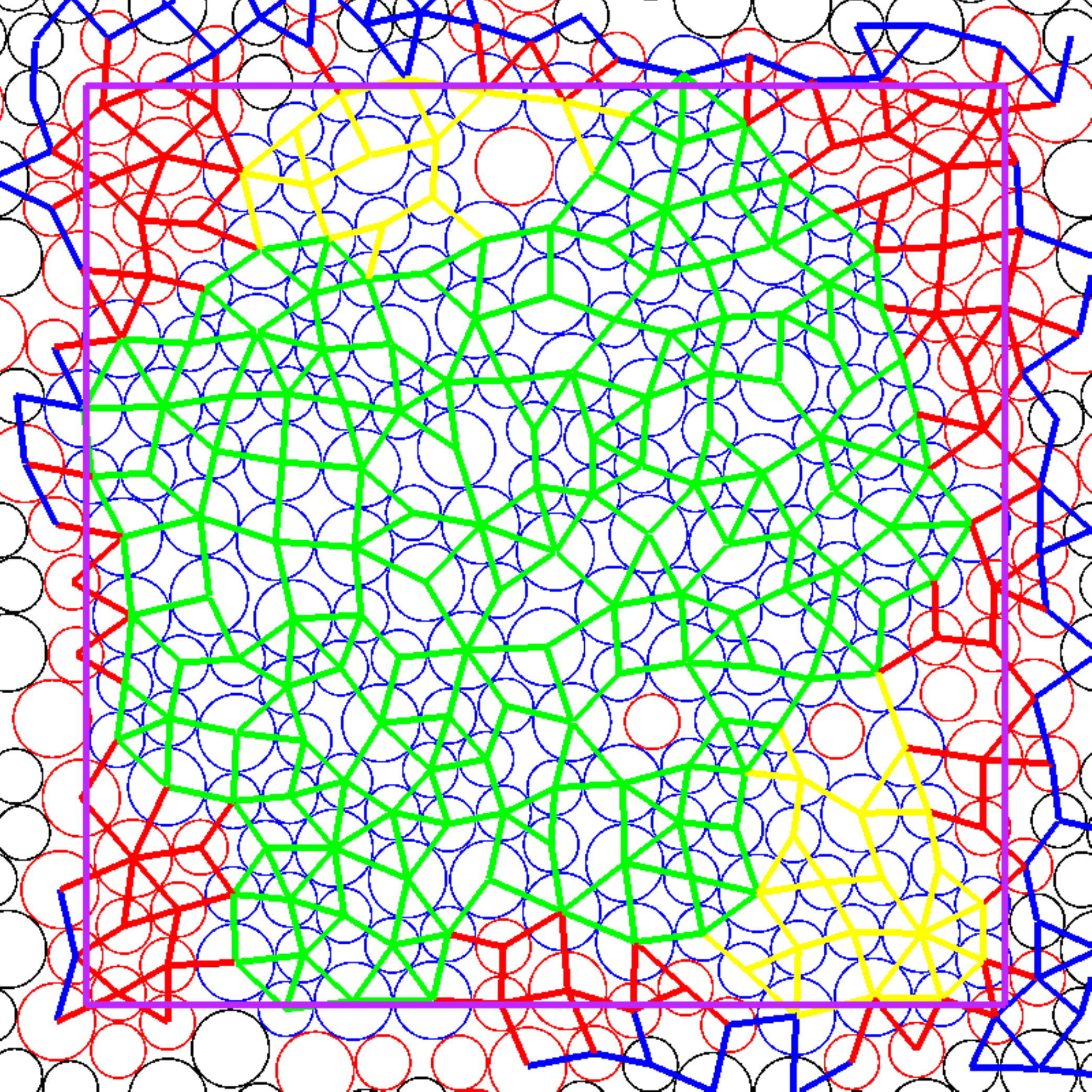}

}\hfill{}\subfigure[]{\includegraphics[scale=0.27]{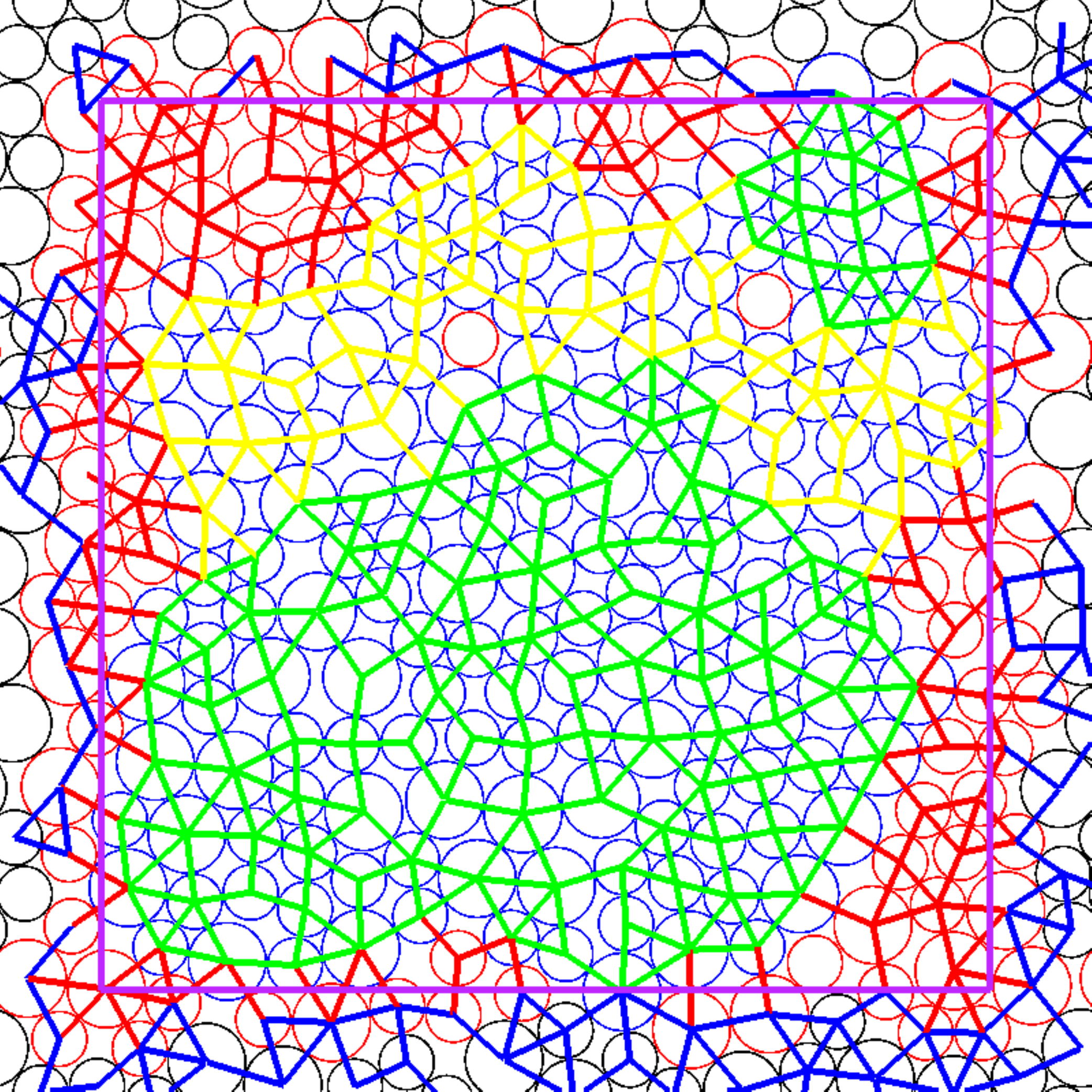}}\caption{a.) Isostatic clusters are not always connected. Here there are two
separate isostatic clusters. b.) On the other hand, underdetermined
clusters, the green regions, are not always connected either. For
each independent green cluster, there is an additional set of constraints
from the stress tensor.}
\end{figure}

While the local corrections to the bulk-surface argument are the most
important contributions to the bulk-surface argument, they are not
the only corrections. A second important correction comes from isostatic
clusters of grains which are embedded inside the boundary of size
$R$. Generally, there are sometimes one (or more) connected sets
of grains within the boundary who's ME equations together define a
linear system which completely determines the corresponding set of
force variables. This set of grains is not necessarily the entire
set of grains within the subregion (if it was, the nullity would be
zero). In addition, linear subsystems embedded within these isostatic
clusters are typically underdetermined, meaning that the linear subsystems
defined by these clusters cannot be reduced to local effects like
those due to the corrections discussed previously. 

\begin{figure}
\subfigure[]{\includegraphics[scale=0.53]{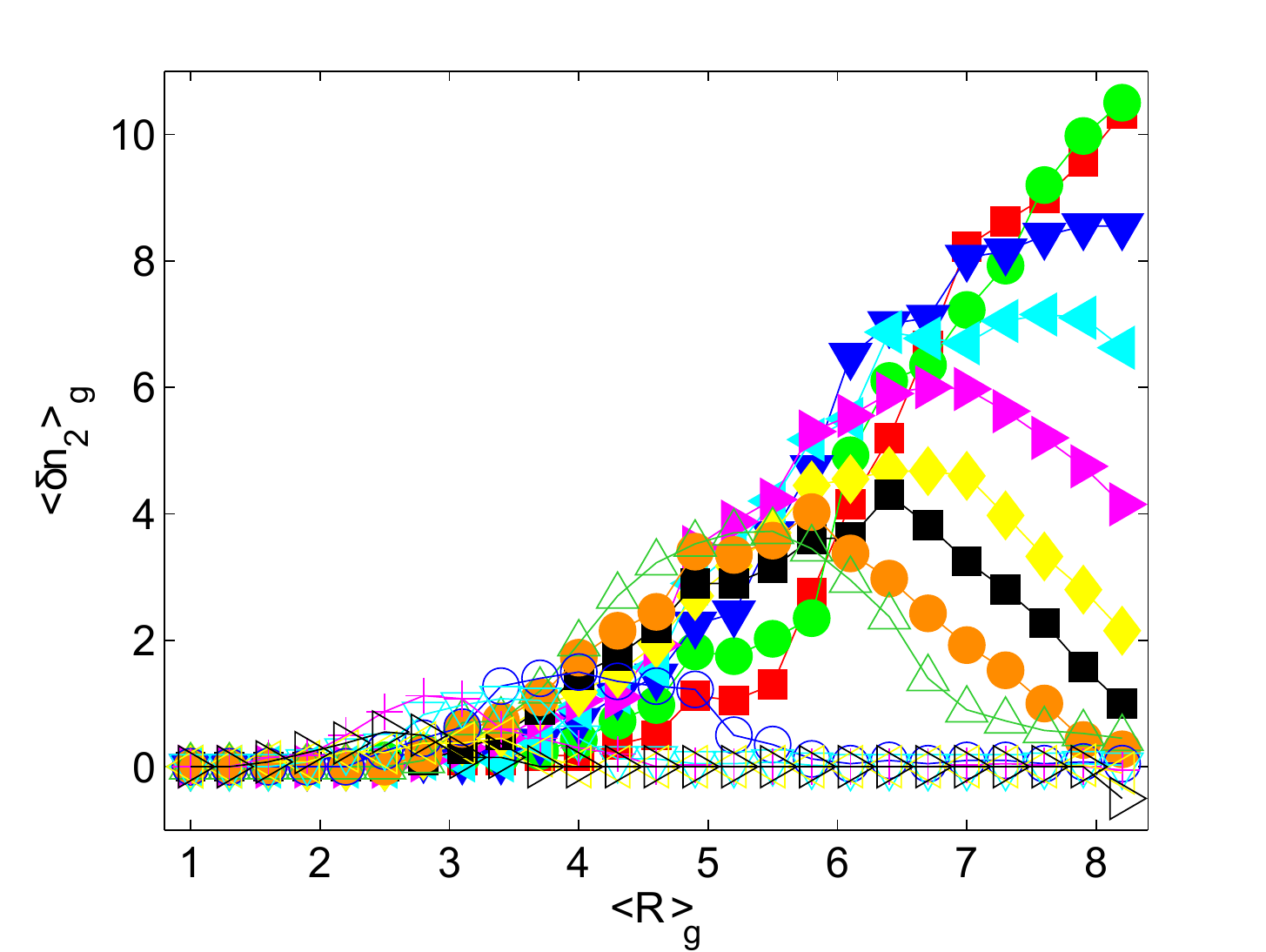}}\hfill{}\subfigure[]{\includegraphics[scale=0.53]{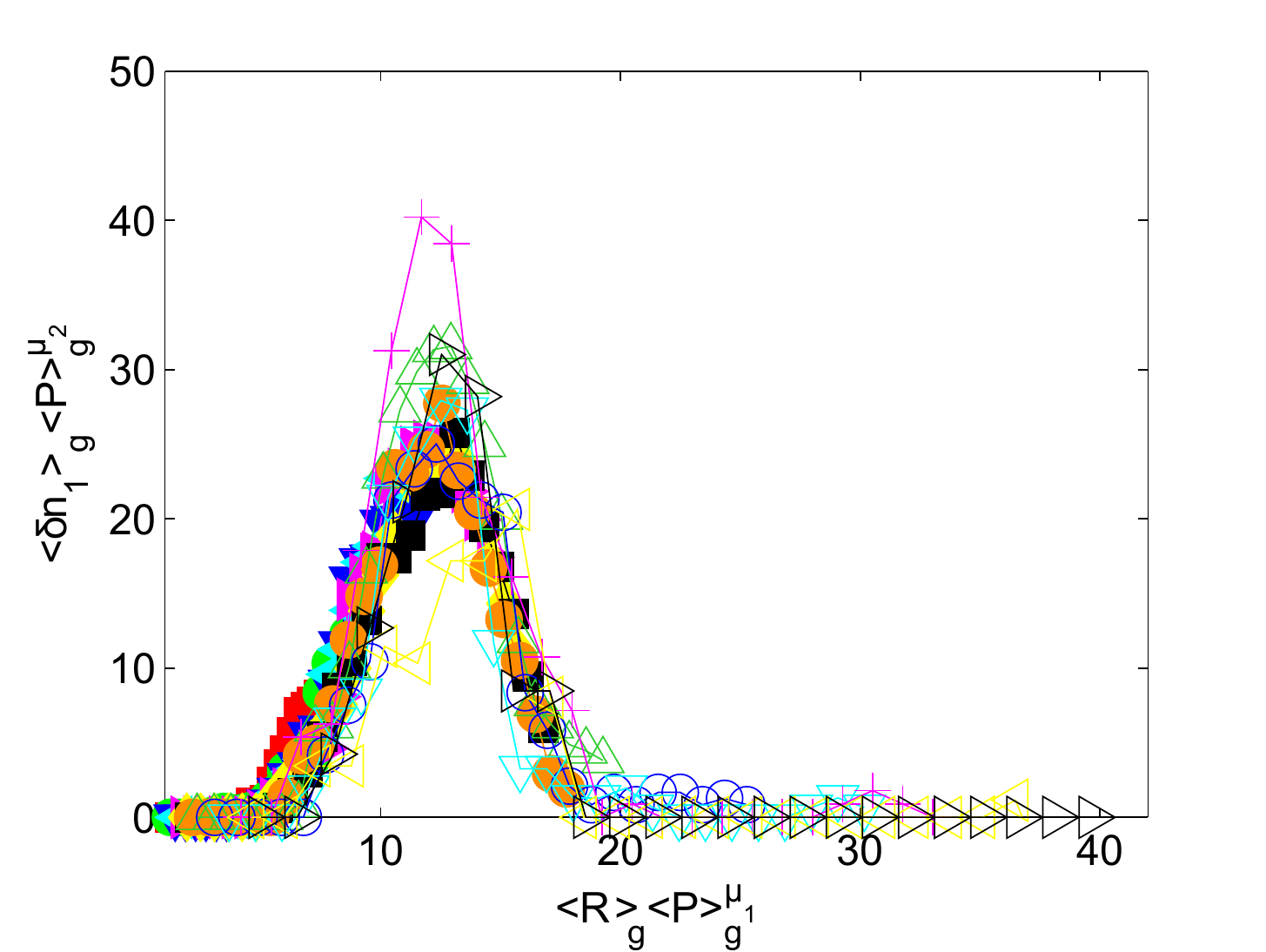}

}\caption{In figure a.) the isostatic cluster correction $\delta n_{2}$, which
is the difference between the number of ME equations and the number
of determined contacts, associated with one of the isostatic (yellow)
clusters. b.) a scaling does exist to collapse this data well; $\mu_{1}=0.4$
and $\mu_{2}=1$. Symbol colors are consistent with those used in figure \ref{fig:C_rho_main_results}.\label{fig:cluster_stats_and_scaling}}
\end{figure}

In addition to having a $P$ dependence, these cluster contributions
to the bulk-surface argument are not linear in $R$, but do exhibit
scaling with the pressure (figure \ref{fig:cluster_stats_and_scaling}). 

\subsection*{Multiple independent clusters\label{sec:Multiple-independent-clusters}}

The combination of locally determined force variables and isostatic
clusters occasionally leads to a fracturing of the subregion defined
by $R$ into multiple underdetermined clusters (i.e., clusters
of force variables which can be changed while satisfying ME). A third,
and less important, correction results from the number of distinct
underdetermined clusters. To be clear, these clusters are $\emph{not}$
the yellow isostatic clusters. For each cluster of underdetermined
forces, the set of determined boundary forces defines a local stress
tensor. Because this stress tensor is fixed as long as the boundary
forces cannot change, the three components (in 2D) of the stress tensor
provide additional relationships between the otherwise independent
equations of ME which make up the linear system associated with the
underdetermined cluster, reducing the total number of independent
equations by three. These fractured subregions do not occur frequently
enough to be characterized statistically in a quantitative way, but
they clearly become more common as the pressure is lowered. 

A simple algorithm is employed to count the number of independent clusters.
First, the algorithm chooses a grain which is undetermined (meaning
a grain whose ME equations are not involved in either a locally determined
or an isostatic cluster). Then, it builds a set of grains out of this
first grain's neighbors, and repeats this process with the new grains
until the set includes the neighbors of each other grain in the set
or a boundary grain, which has all determined forces. Then the algorithm
restarts with a grain not included in the previous set. It continues
to build these new sets until all of the undetermined grains have
been exhausted. The number of times the algorithm restarts is the
number of independent clusters.

\begin{figure}[t]
\subfigure[]{\includegraphics[scale=0.53]{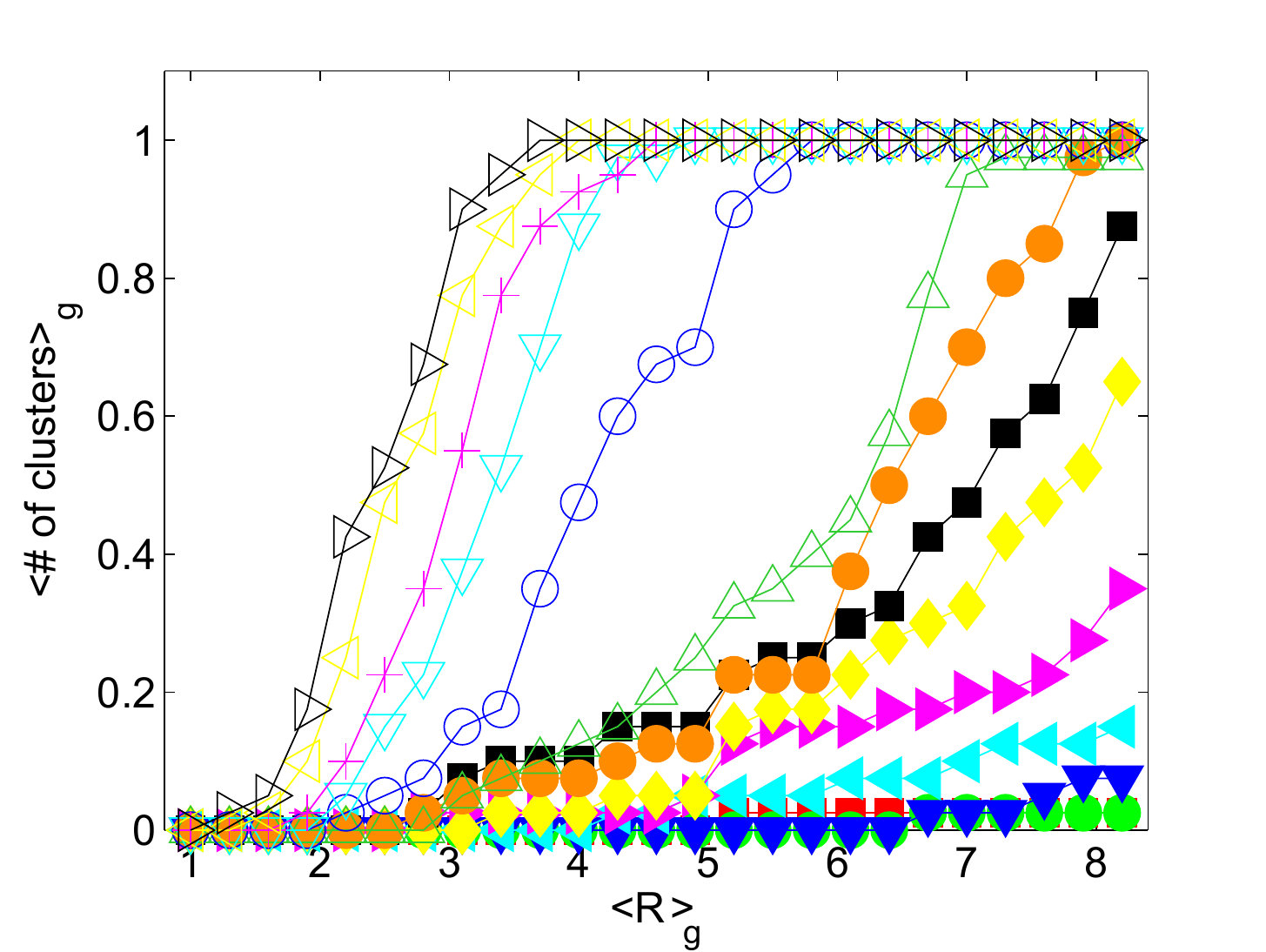}}\hfill{}\subfigure[]{\includegraphics[scale=0.53]{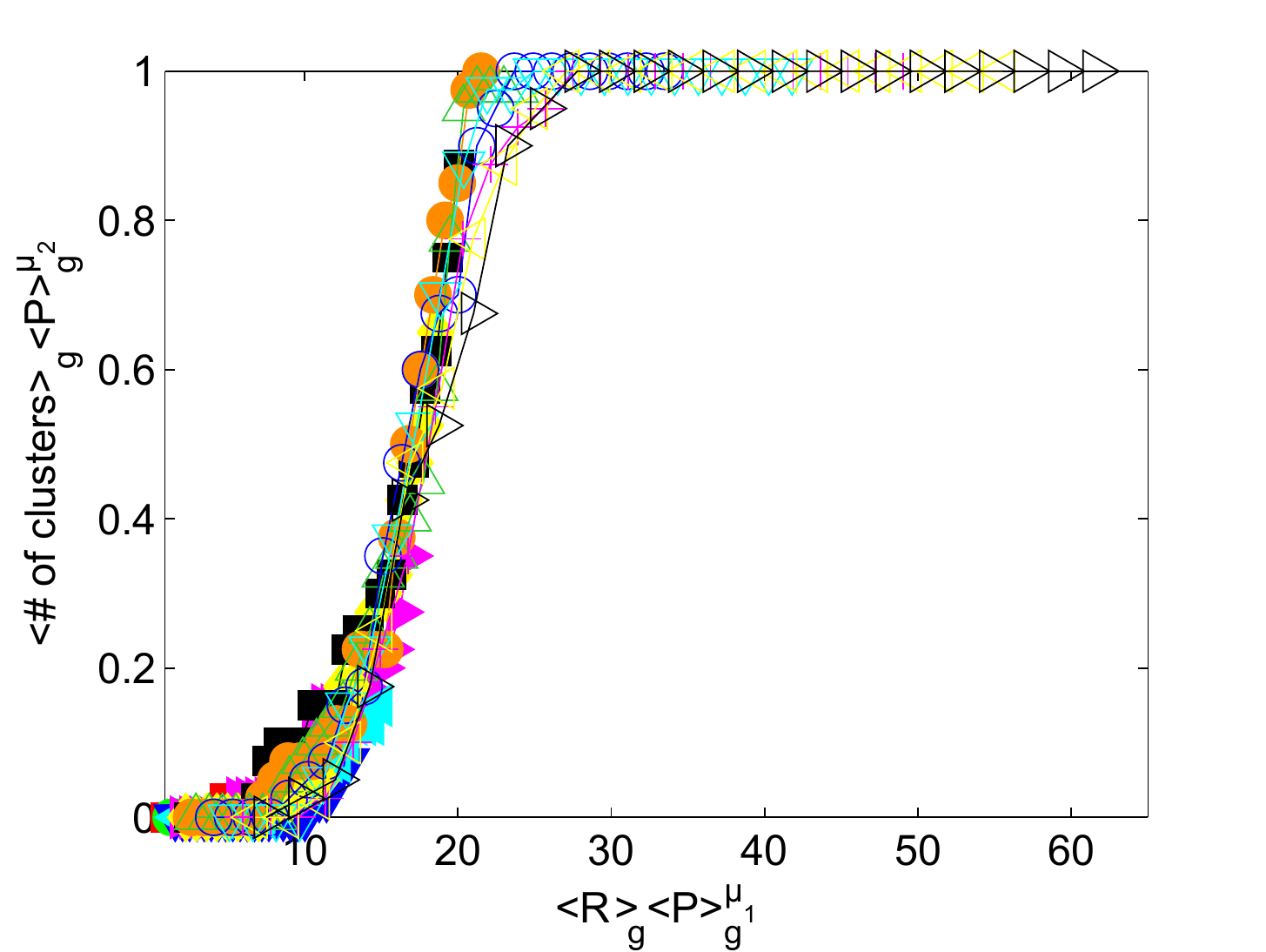}

}\caption{The number of clusters rarely becomes greater than 1 for this system
size, and so the 2-cluster packings are averaged out during geometry-averaging.
Figure a.) shows the number of clusters, while figure b.) shows a
simple scaling of the horizontal axis with $\mu_{1}=0.5$ and $\mu_{2}=0$.
This is to be expected, since the number of clusters goes from 0 to
1 as soon as $\delta n$ becomes non-zero. The scaling of the horizontal
axis reflects the scaling with $R_{0}$. Symbol colors are consistent with those used in figure \ref{fig:C_rho_main_results}.}
\end{figure}

\subsection*{Locally crystalline}

\begin{figure}[t]
\subfigure[]{\includegraphics[scale=0.27]{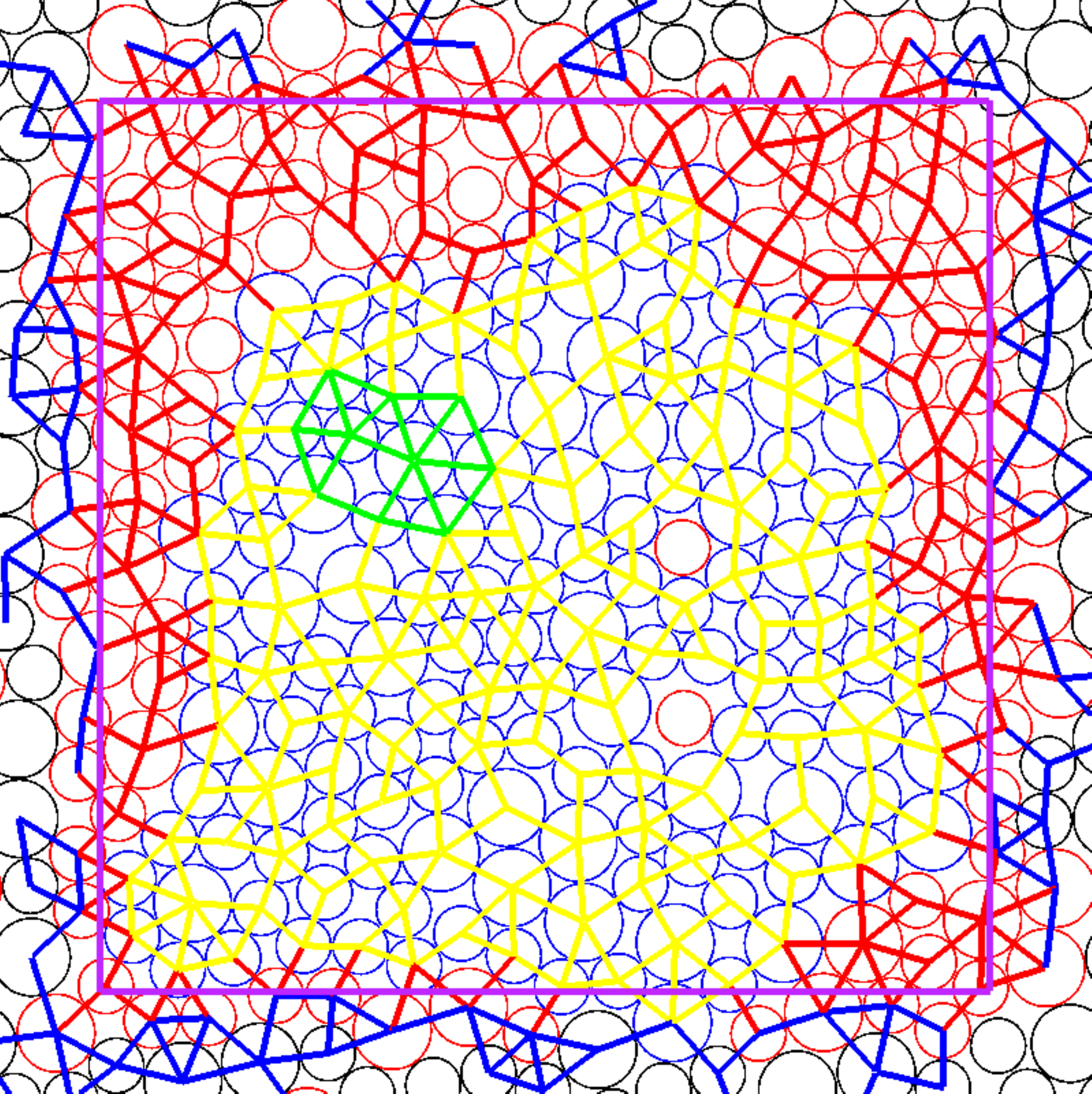}}\hfill{}\subfigure[]{\includegraphics[scale=0.27]{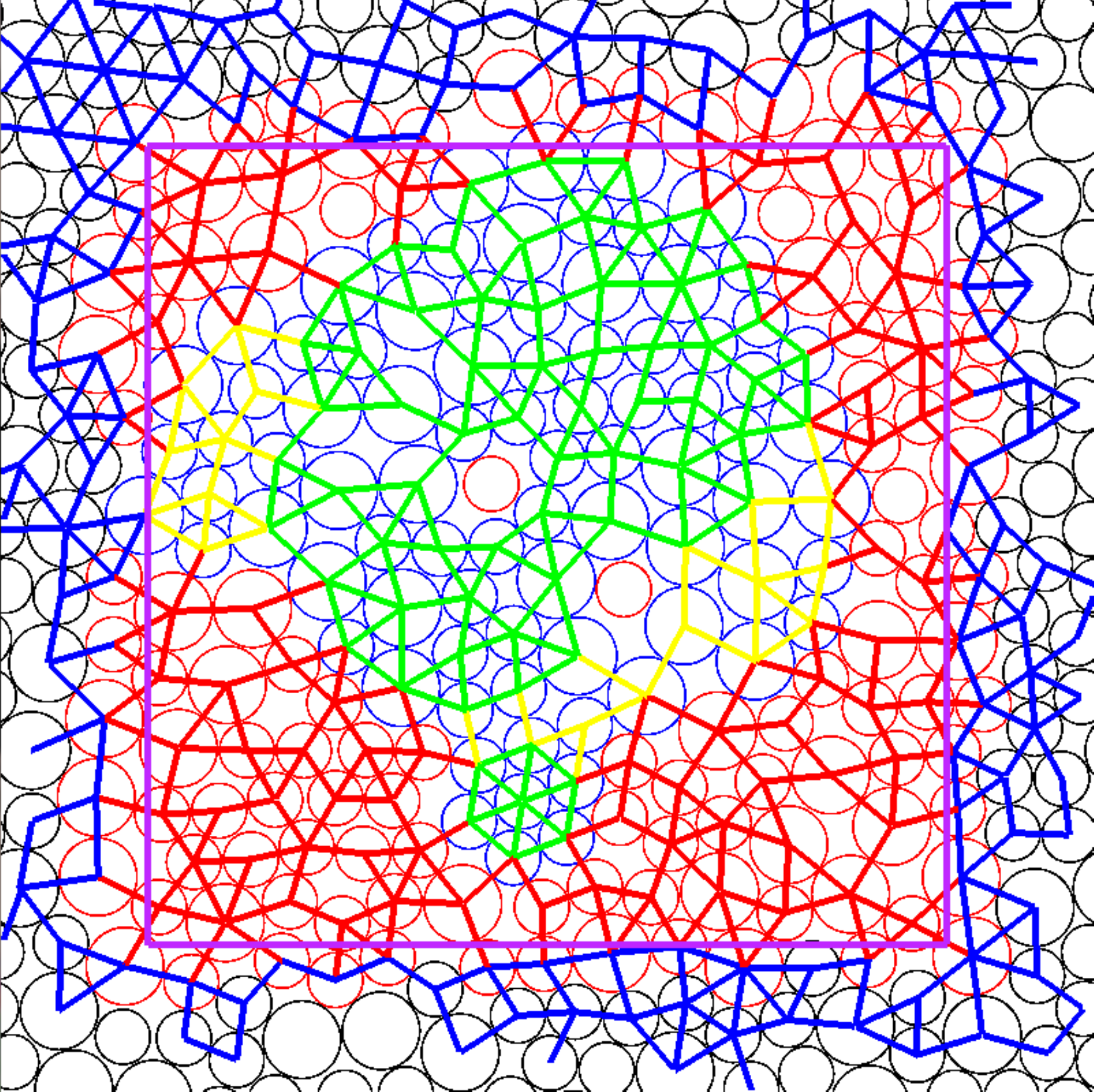}

}\caption{Figure a.) shows an extremely localized underdetermined cluster (green).
The isostatic cluster has propagated far into the bulk of the packing,
leaving only a small cluster to fluctuate. In the bottom of figure
b.), there is a crystalline cluster, separated from the main cluster
by two isolated yellow bonds. At first, it may seem odd that those
yellow bonds are considered determined. It turns out, though, that
these yellow bonds, along with the crystalline cluster, form an isostatic
cluster. Because of its symmetry, however, the crystalline cluster
can still fluctuate according to wheel moves. For this packing, the
null vectors have been analyzed, and the crystalline cluster is in
fact a wheel move and no other types of fluctuations are allowed.
\label{fig:wheel-move cluster}}
\end{figure}

Finally, even after the previously discussed corrections to be bulk-surface
argument are taken into account, there are very rarely special cases
where a single force variable shared by two grains can be determined
(in the sense that it is already known), even though none of the other
force variables associated with those grains are. This may seem counter
intuitive. All of the previously discussed phenomena involve a boundary
effect where a grain whose force variables are all determined is in
contact with a grain whose force variables are not all determined.
At first, it might seem that if all of a grain's contact force variables
can change but one, there is no reason why that one should not fluctuate
as well. In the special case of crystalline order, though, we know
from experience with the {}``wheel move'' that the ME equations
are not all independent, leading to additional {}``breathing'' modes.
Take, for instance, the localized wheel move type cluster in the bottom
of figure \ref{fig:wheel-move cluster}. There are 14 ME equations
associated with this cluster, and including the two yellow links which
point upward (and are shared between grains which also contain undetermined
force variables, which is what makes this example unique) there are
14 force variables. Assuming the two vertical links do indeed correspond
to determined variables, the wheel move still allows for the 12 internal
green links to fluctuate without changing any of the other forces,
simply because of the added symmetry of the triangular lattice. Or,
to put it another way, because the ME equations associated with a
crystalline structure are not all independent, some forcs must be
determined and the SVD essentially {}``chooses'' to localize the
basis on the crystalline structure and never allow the boundary force
variables to fluctuate. Because the wheel move is in some sense a
different underdetermined cluster, it does count towards the total
number of underdetermined clusters. But, since it is not separated
from the main cluster by determined grains, only some particularly
fortuitous determined forces, it is not detected by the cluster detection
discussed in \ref{sec:Multiple-independent-clusters}.

It is not enough to simply have crystalline structure; the crystalline
structure must be surrounded by fixed forces as well, or there would
not be an independent set of stress tensor constraints on the cluster
and so would be indistinguishable from the main cluster. As a result,
these wheel move clusters appear to be extremely rare; too uncommon, in fact, to allow for meaningful statistics.

\subsection*{Net Corrections}

\begin{figure}[t]
\subfigure[]{\includegraphics[scale=0.53]{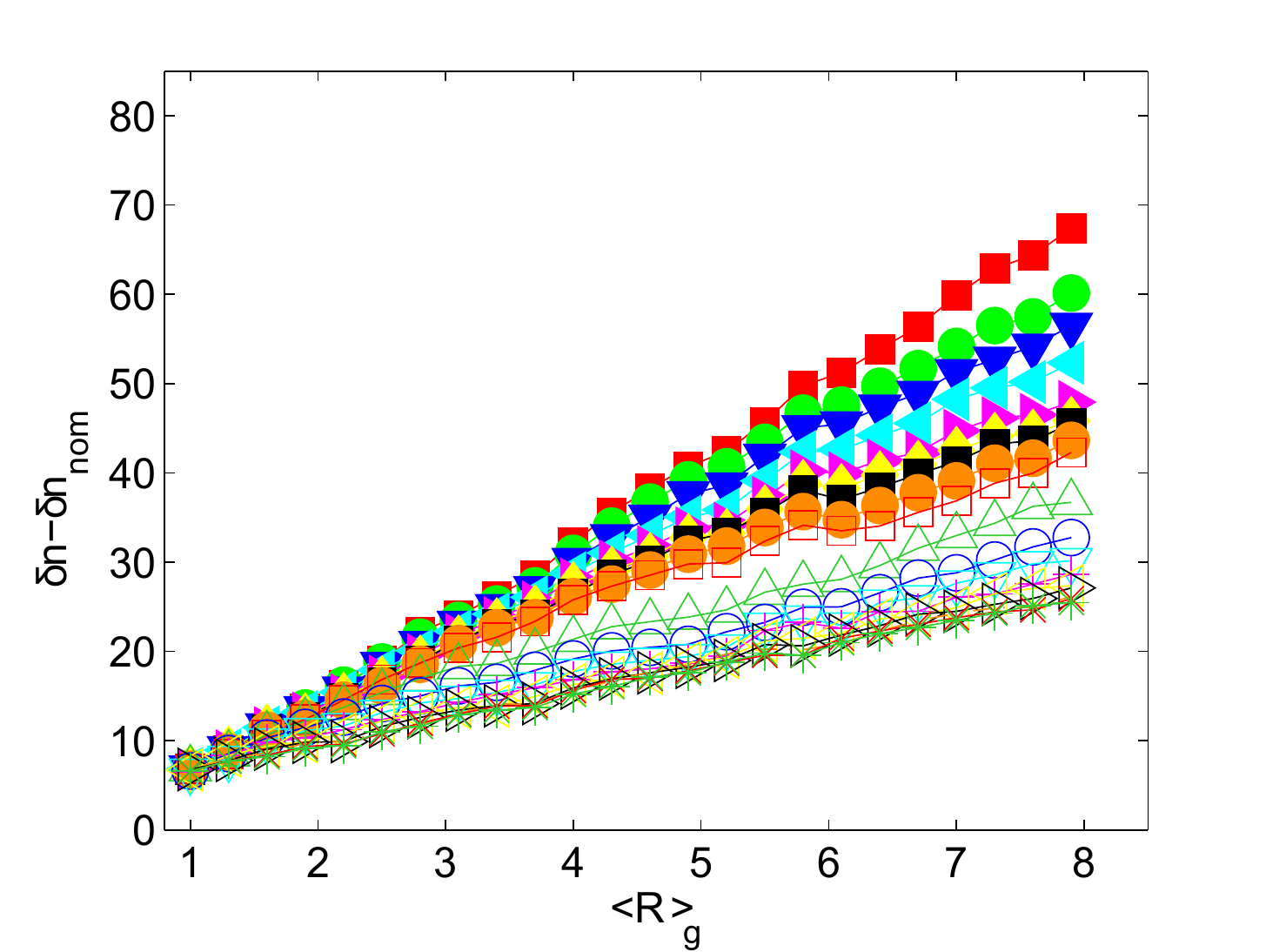}

}\hfill{}\subfigure[]{\includegraphics[scale=0.53]{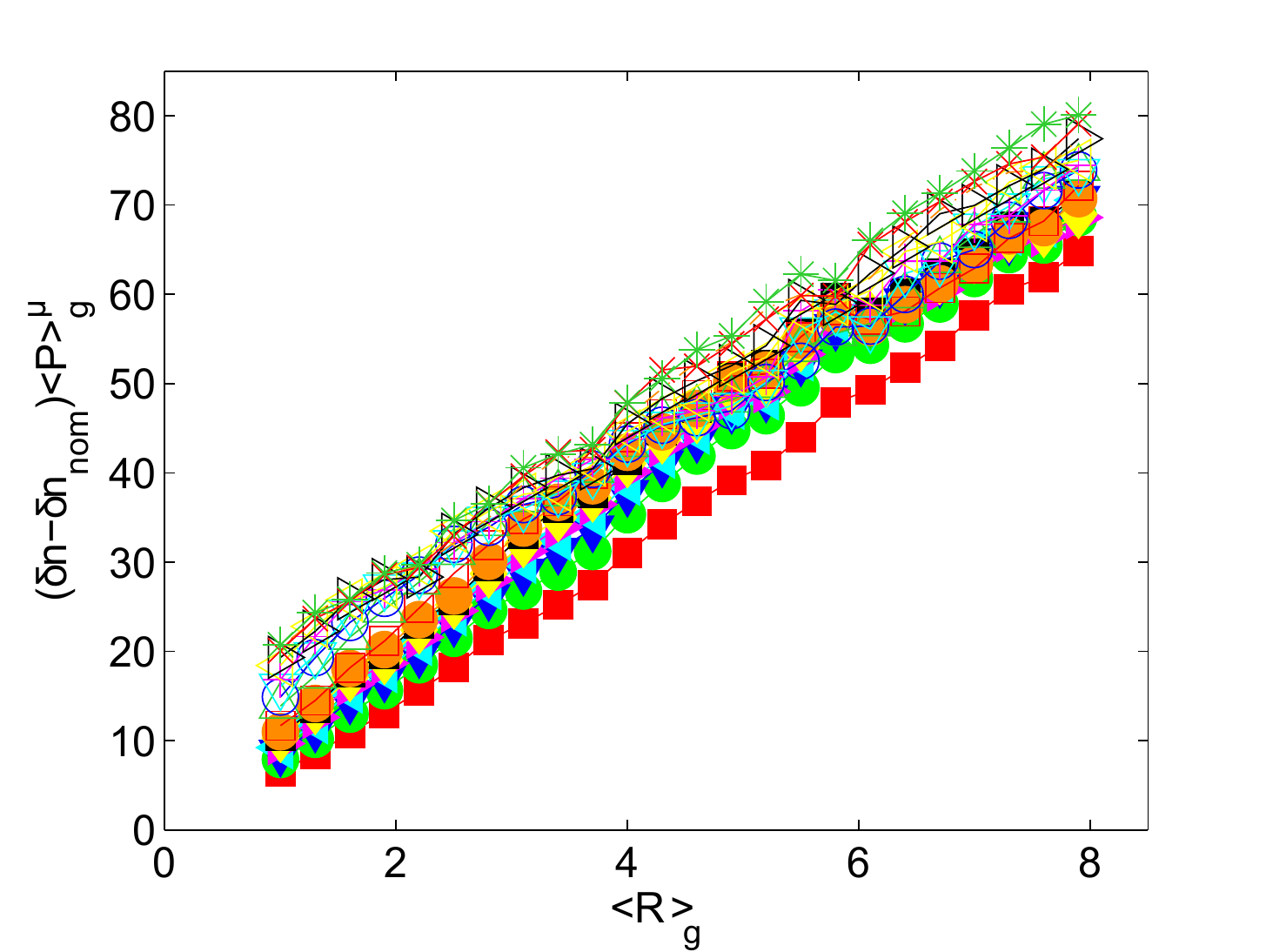}

}

\caption{a.) The total correction to the nominal nullity is shown, which is
linear in $R$ but has a $P$ dependence. b.) The correction scales
with $P$, with an exponent of $\mu=0.25$. Symbol colors are consistent with those used in figure \ref{fig:C_rho_main_results}.\label{fig:The-total-correction}}

\end{figure}

In figure \ref{fig:The-total-correction} the difference between $\delta n$
and $\delta n_{nom}$ is shown. The difference $\delta n-\delta n_{nom}$
represents the total correction to the bulk-surface argument, which
takes the form $\delta n_{nom}=aR^{2}P^{1/2}-bR$ for constant
$a$ and $b$ (the effects of slight variations in  packing fraction are
being ignored). The correction $\gamma(R,P)=\delta n-\delta n_{nom}$
amounts to an additional term in the bulk-surface argument: $\delta n=aR^{2}P^{1/2}-bR+\gamma(R,P)$.
If this term were to be either linear and independent of pressure,
or quadratic and dependent on the square root of pressure, the mean-field
exponent $\nu=0.5$ would be recovered. As seen in figure \ref{fig:The-total-correction},
neither is the case. $\gamma(R,P)$ is linear in $R$, with a slope
of $P^{-1/4}$, and a pressure dependent vertical intercept $Y\left(P\right)$.
The bulk-surface argument, then, has a form $\delta n=aR^{2}P^{1/2}-(b-P^{-1/4})R+Y(P)$,
which does not have a root of $R_{0}\propto P^{-1/2}$.  Even though there is no theory predicting this form for $\delta n$, it's worth noting that RFOT would predict a free energy as a bulk-surface competition with a temperature dependent surface tension \cite{BiroliNPhys} (in analogy to the pressure dependent coefficient of the linear term found here).

\begin{figure}[t]
\begin{centering}
\subfigure[]{\includegraphics[scale=0.18]{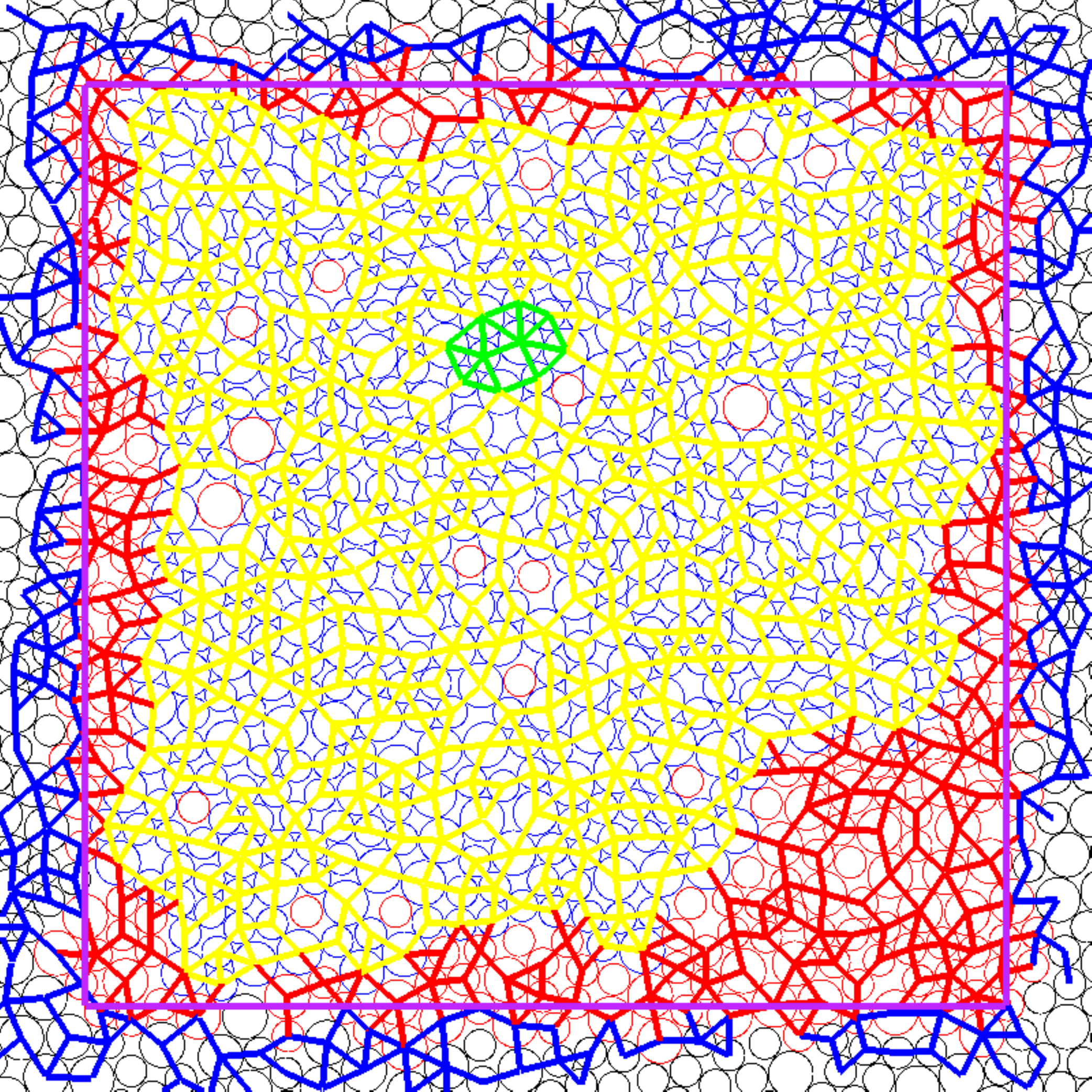}}\qquad{}\subfigure[]{\includegraphics[scale=0.18]{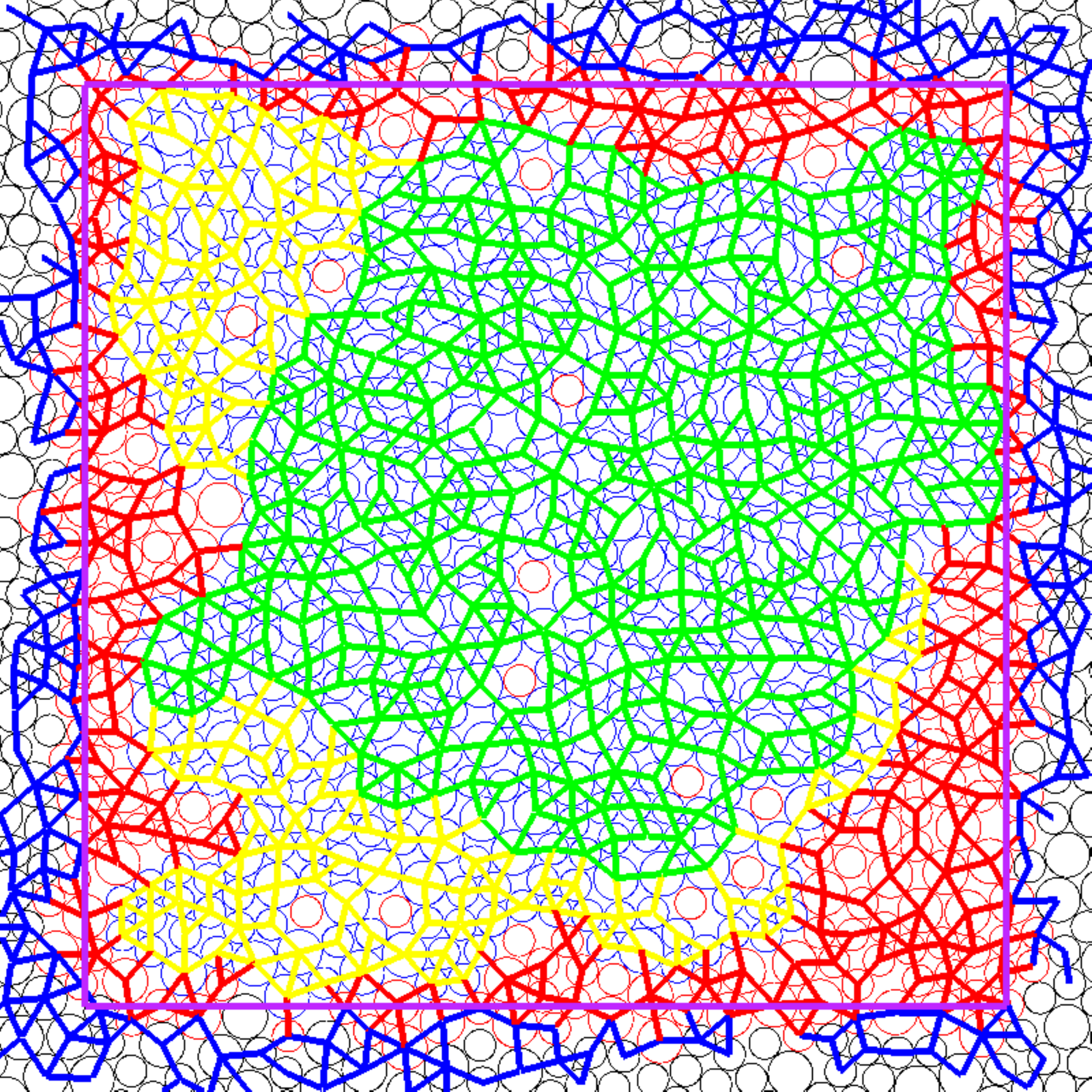}}
\par\end{centering}

\begin{centering}
\subfigure[]{\includegraphics[scale=0.18]{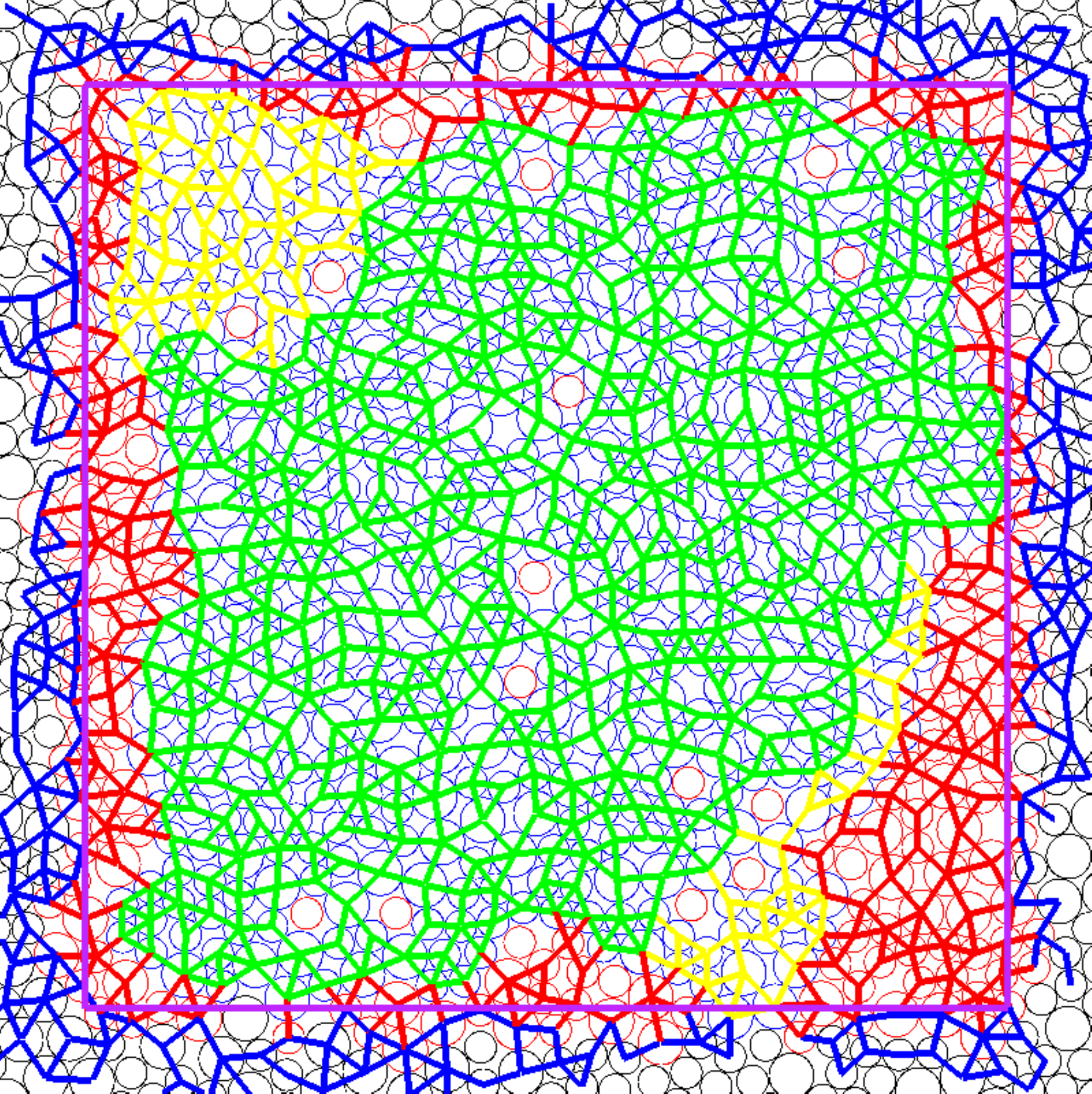}}\qquad{}\subfigure[]{\includegraphics[scale=0.18]{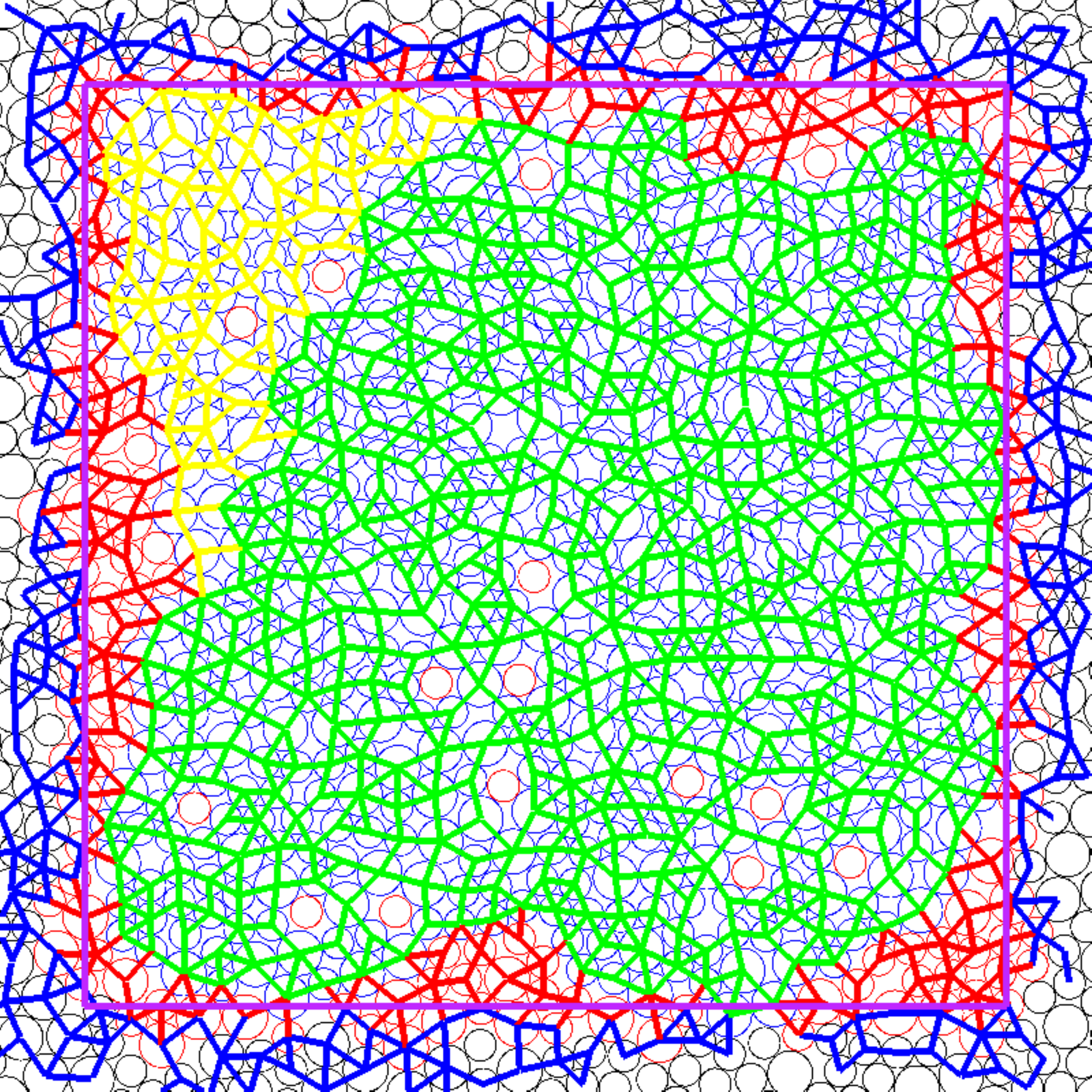}}
\par\end{centering}

\begin{centering}
\subfigure[]{\includegraphics[scale=0.18]{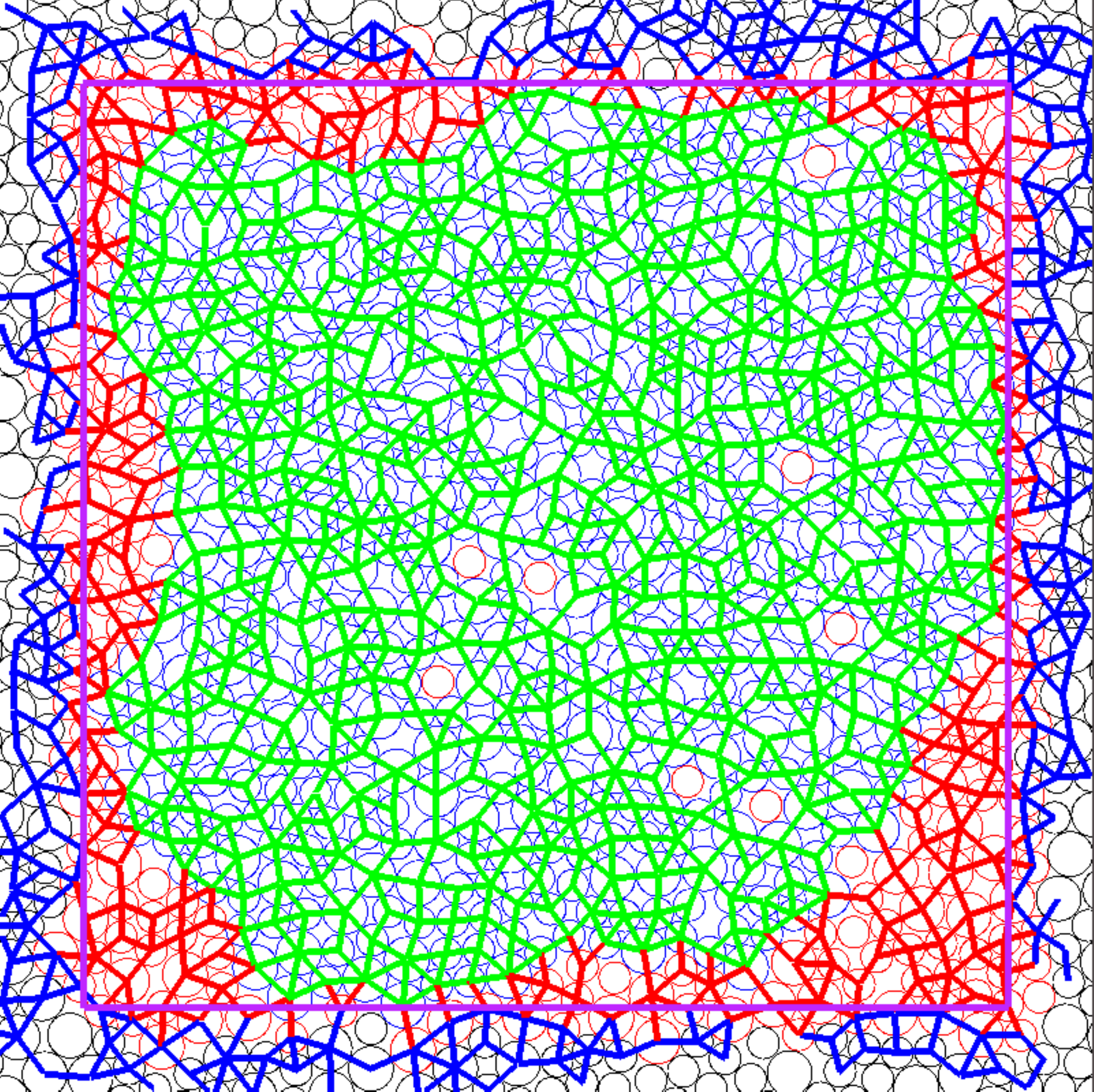}}
\par\end{centering}

\caption{A single 900 grain trajectory is shown here. $R$ is kept fixed
as $\delta\phi$ is increased. a.) the pressure is the lowest, and
the isostatic cluster is nearly the size of the system. Also, the
red regions propagate the furthest into the bulk. b.) $\delta\phi$
has been doubled, from $0.001$ to $0.002$, and the isostatic cluster
has fragmented into two clusters. c.) the isostatic clusters have
shrunk considerably as the undetermined cluster grows. d.) One of
the isostatic clusters has disappeared; on the other hand, the remaining
cluster has grown somewhat. e.) at $\delta\phi=0.007$, the isostatic
clusters have disappeared entirely. \label{fig:sample_decompression_traj}}

\end{figure}

This section is concluded with an illustration of the clusters that
emerge in a single packing as the pressure is lowered (figure \ref{fig:sample_decompression_traj}).
The figure caption addresses the particular characteristics of each
step in the decompression. 

\section{Discussion \label{sec:Discussion}}

This article has detailed a new measure of correlations that arise
in jammed granular systems. The model system is the force network ensemble constructed on geometries of compressible frictionless
disk packings.  The correlation function that has been introduced probes the effects of boundaries on mechanically stable packings, and 
identifies a length scale that diverges as the packing unjams. A description
of the unjamming transition as a critical point is developed based
on the configurational entropy of equivalent mechanically stable force
configurations.

The configurational entropy is itself based on a solution space picture
where the mechanically stable force networks fill a configuration
space that shrinks in size and dimension as the packings unjam. The
length scale is shown to result from this entropy loss. A comparison
of this length scale to the mean-field predictions of a popular bulk-surface
argument has shown that the true exponent associated with the diverging
length scale deviates from the mean-field prediction of 0.5. Analysis
of the configuration space shows that this deviation is the result
of the failure of the bulk-surface argument itself. Analysis of microscopic
properties of the packings show that the failure of the bulk-surface
argument is the result of additional terms which when taken together
are linear in $R$ but depend on pressure with a novel exponent.
The bulk-surface argument relies on all ME constraints to be independent
of each other, an assumption which is shown to fail in 2D disk packings
close to the unjamming transition. 

\begin{figure}[t]
\begin{centering}
\includegraphics[scale=0.53]{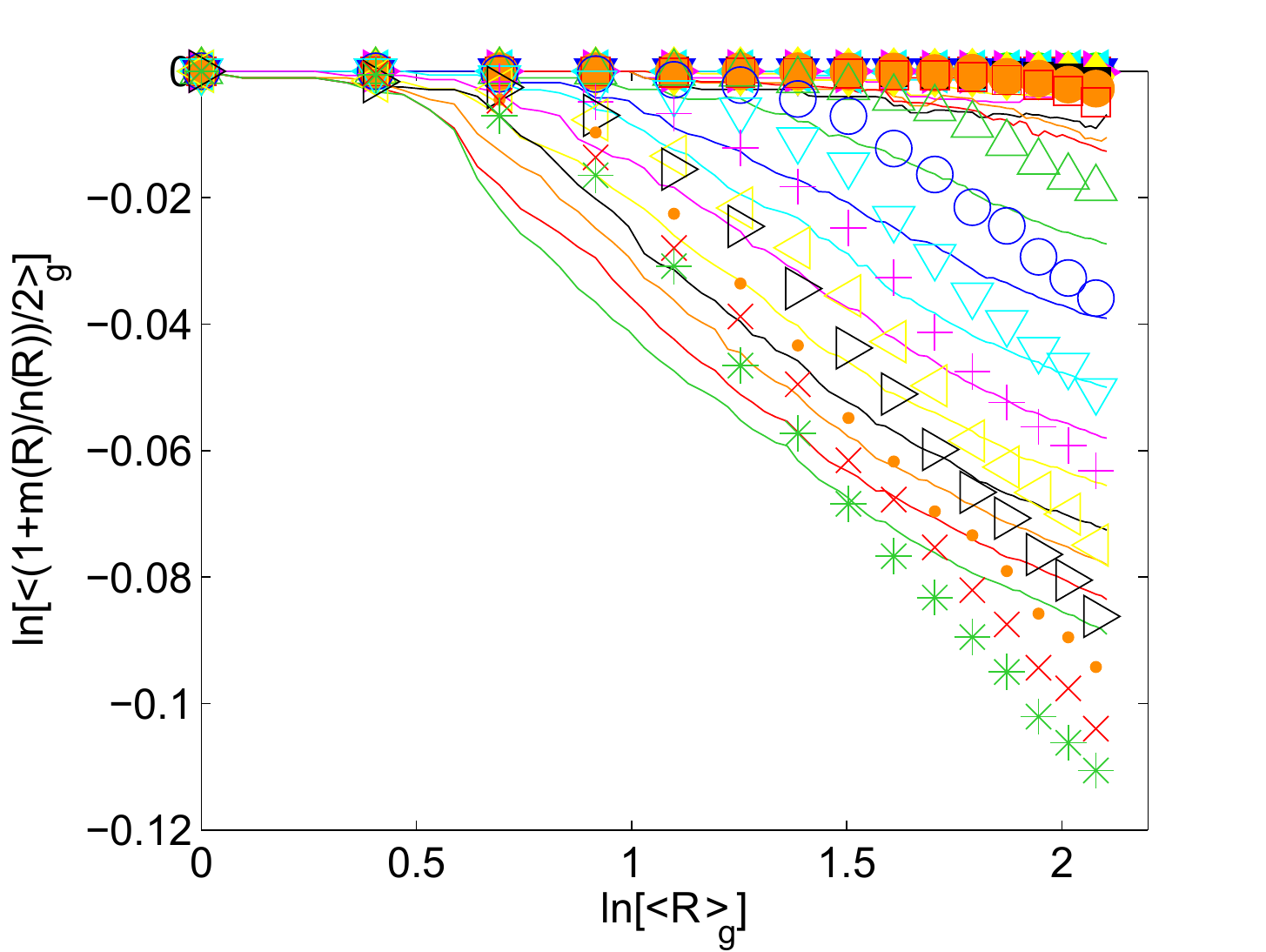}
\par\end{centering}

\caption{The results of Eq.\ref{eq:model_C} are shown here, with numerical
results for the corrected values of $m\left(R\right)$ and $n\left(R\right)$
used. $\left\langle \left\langle C\left(R\right)\right\rangle \right\rangle _{g}$
is shown for reference. There is qualitative agreement with the model
and $\left\langle \left\langle C\left(R\right)\right\rangle \right\rangle _{g}$,
but with logarithmic scaling of the axes, the model does not seem
to capture the power-law form of the tail. Symbol colors are consistent with those used in figure \ref{fig:C_rho_main_results}.\label{fig:C_model_results}}

\end{figure}

The correlation function $C\left(R\right)$ is measured directly
in numerics, and exhibits two regimes. The first is a plateau at $C\left(R<R_{0}\right)=1$,
and the crossover between this plateau and the tail of $C\left(R\right)$
identifies the length scale. The tail $C\left(R>R_{0}\right)$
is also described by a model based on properties of the solution space.
Here we revisit Eq.\ref{eq:model_C}. One important result of section
\ref{sec:Corrections-to-l*} is that a simple counting of all contacts
and grains within the boundary of size $R$ is not sufficient to
characterize the number of free variables and constraints, since some
constraints are not independent. Instead, an analysis of the null
space of $A\left(R\right)$ is used to identify the true number
of free variables (the nullity) and algorithms are developed to find
the independent constraints. It is these quantities which must be
inserted into Eq.\ref{eq:model_C}, and so it is only at this point
that Eq.\ref{eq:model_C} can be compared to the direct measurements
of $\left\langle \left\langle C\left(R\right)\right\rangle \right\rangle _{g}$
(see figure \ref{fig:C_model_results}) . Eq.\ref{eq:model_C} is
in rough agreement with the measurements of $\left\langle \left\langle C\left(R\right)\right\rangle \right\rangle _{g}$,
in that it captures exactly the same length scale (by construction),
and decays similarly to $\left\langle \left\langle C\left(R\right)\right\rangle \right\rangle _{g}$.
But, it fails to capture the power-law form of the tail, or agree
quantitatively with the tail. 

There is still a great deal of work left to do. For instance, while
many microscopic sources of corrections to the mean-field bulk-surface
argument have been identified, there is no proof that they have all
been found. Are there other types of dependencies between constraints,
for instance in much larger systems or much closer to the critical
point, that arise? What is the form of $Y\left(P\right)$, and how
does it affect the exponent $\nu$? Since the constraints become dependenft,
this suggests that there is some loss of randomness or disorder in
the packings as the critical point is approached. One question that
arises is whether or not an order parameter can be constructed based
on these observations. In addition, the question of the importancefa
of dimensionality is not addressed, since only 2D systems have been
studied. 

The mean-field corrections that are explored in section \ref{sec:Corrections-to-l*}
identify structures within the packing that grow larger as the unjamming
transition is approached. The immediate result of these observations
is that the mean-field exponents do not apply. What's more, correlations
lead to the emergence of precisely determined contact forces over
larger scales when a packing is closer to unjamming. Is there, for
instance, a renormalization group approach that could predict the observed
exponent?

Also of interest is how grain shape affects the nature of the PTS correlations that arise in jammed packings.  Specifically, we have considered 2D packings of elliptical grains, which are known to be \emph{hypostatic}\cite{DonevEllipses}.  That is to say, elliptical grain packings, even when highly overcompressed, have less contacts than required for rigidity by the isocounting procedure when accounting for additional torque balance constraints.  Nonetheless, elliptical grain packings are found to be mechanically stable.  One possibility is that elliptical grain packings are ordered in a non-obvious way that leads to linear dependencies between ME equations (including torque balance equations).  Through direct numerical analysis we have verified that such packings are in fact {}``critical": we always find exactly the same number of contacts as linearly independent constraints in ellipse packings and so the PTS correlation length as described in this work is the size of the system.  This result holds in every case that we've checked below $[z]=6$, which is the prediction from the isocounting procedure.  At overcompressions resulting in $[z]>6$, elliptical grain packings begin to exhibit underdetermined contact forces (see figure \cite{fig:EllipsePhaseDiagram}).  And so an intriguing question arises: do elliptical grain packings exhibit a line of true critical points from the marginally jammed, hypostatic packings up to $[z]=6$?  Or, is there a more appropriate construction of the PTS correlation function that describes unjamming of elliptical grain packings? 

\begin{figure}[t]
\begin{centering}
\includegraphics[scale=0.43]{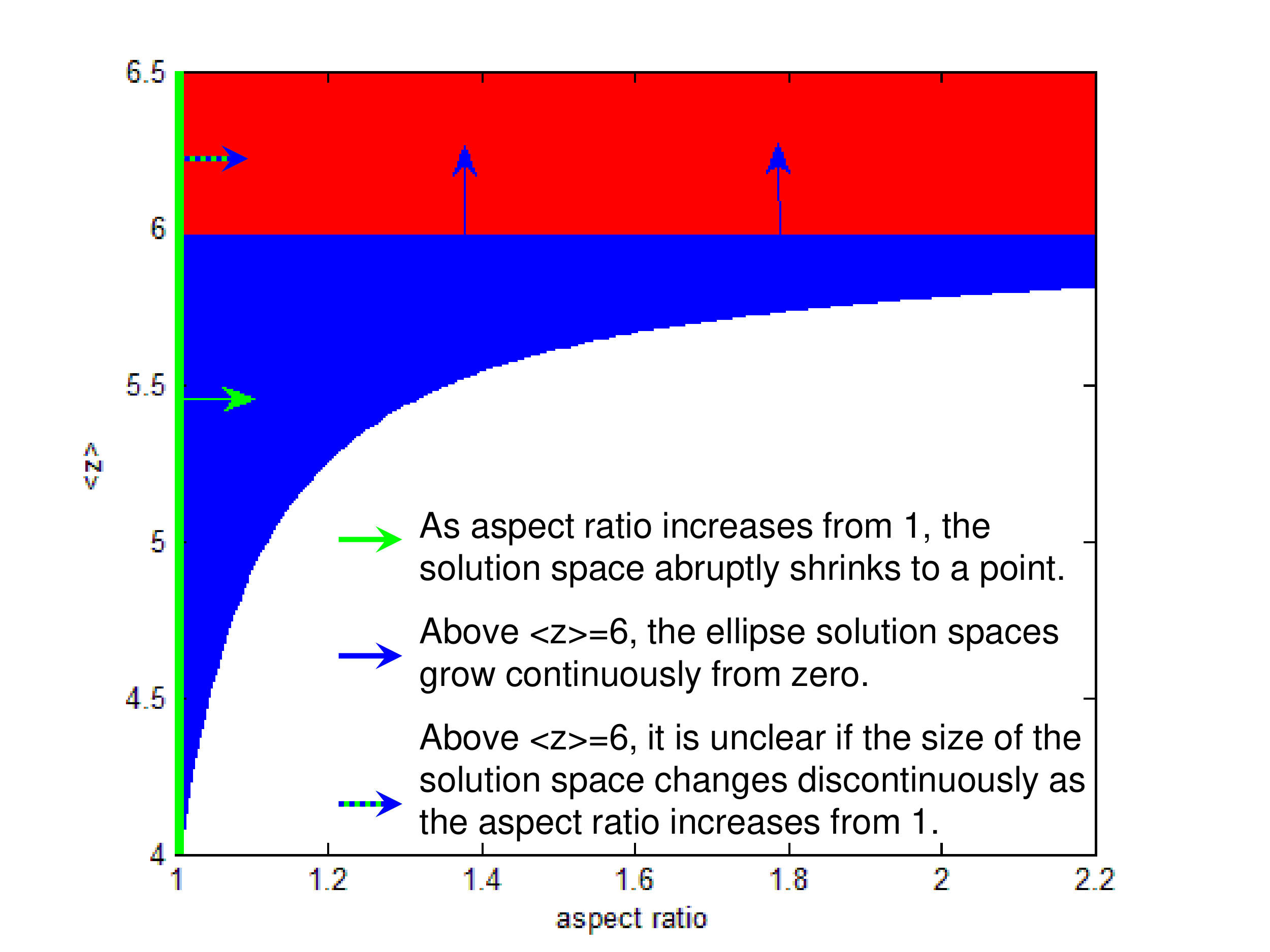}
\par\end{centering}

\caption{A schematic of an elliptical grain phase diagram exhibits several important regions.  The region in white is unjammed, while the vertical green line represents disk packings (aspect ratio 1) which are jammed from $z_0=4$.  In the blue region hypostatic ellipse packings are mechanically stable.  These packings are differentiated from those occupying the red region, which are heavily overcompressed and are mechanically stable, but additionally have non-zero nullity (are underdetermined with respect to contact forces). \label{fig:EllipsePhaseDiagram}}

\end{figure}

Finally, a good deal of theoretical work concerning the solution space
is left undone. Does a convincing argument exist that the solution
space is convex, and if it is not, what are the implications for the
approximation of the solution space as a roughly hyperspherical structure?
There also may be instances where the solution space is not isotropic,
for instance when packings are under shear, and the weights used in
sampling of the solution space may very well be important. This work
always applies a flat measure to the solution space, so that each
force network is equally likely. In RFOT, it is acknowledged that
the free energies of different glass states are not the same, and
so some states are much more likely than others. In fact, the length
scale derived from the PTS correlation function in RFOT relies on
this distribution of free energies, and the length scale is extracted
from the tail of the PTS correlation function rather than from the
plateau. An interesting challenge for the granular PTS correlation
function is to attempt to create a tail that exhibits non power-law
behavior by sampling force networks non-uniformly, and seeing if the
tail exhibits a different length scale. Are there physically realistic
sampling biases that could be applied? For instance, it may be that
applying a small amount shear to a packing can be captured by a non-uniform
sampling of force networks.

\ack
This work was supported by NSF-DMR0905880, and has benefited from the facilities and staff of the Yale University High Performance Computing Center and NSF CNS-0821132. We acknowledge
useful discussions with S. Franz, G. Biroli, Dapeng Bi, N. Menon and C. S. O'Hern, and with participants at the 2010 Les Houches Winter School. BC acknowledges the Aspen Center for
Physics, and the Kavli Institute for Theoretical Physics where some of this work was done.

\section*{References}
\bibliographystyle{unsrt} 
\bibliography{BibThesis}

\end{document}